\documentclass[letterpaper,12pt]{extreport}
\usepackage{times}
\usepackage{amsmath,amssymb,braket,slashed,mathrsfs}
\usepackage{epsfig}
\usepackage{color}
\usepackage{cite}
\usepackage{setspace}
\usepackage{wasysym}

\usepackage[top=15mm,bottom=20mm,left=15mm,right=15mm]{geometry}

\usepackage{url}

\usepackage{titlesec}
\titleformat{\chapter}[display]
  {\Large\bf}
  {}{20pt}{\Large}

\usepackage{rotating}

\def\be{\begin{equation}}
\def\ee{\end{equation}}
\def\bea{\begin{eqnarray}}
\def\eea{\end{eqnarray}}

\def\reff#1{\ref{#1}}
\def\labell#1{\label{#1}}

\def\eq#1{Eq.~(\reff{#1})}
\def\eqs#1#2{Eqs.~(\reff{#1}) and (\reff{#2})}

\def\eqsmany#1#2{Eqs.~(\reff{#1})-(\reff{#2})}
\def\sec#1{Sec.~\reff{#1}}
\def\secs#1#2{Secs.~\reff{#1} and \reff{#2}}
\def\secss#1#2#3{Secs.~\reff{#1}, \reff{#2} and \reff{#3}}
\def\fig#1{Fig.~\reff{#1}}

\def\tab#1{Tab.~\reff{#1}}
\def\tabs#1#2{Tabs.~\reff{#1} and \reff{#2}}

\newcommand{\lsim}{ {\
\lower-1.2pt\vbox{\hbox{\rlap{$<$}\lower5pt\vbox{\hbox{$\sim$}}}}\ } }
\newcommand{\gsim}{ {\
\lower-1.2pt\vbox{\hbox{\rlap{$>$}\lower5pt\vbox{\hbox{$\sim$}}}}\ } }

\def\er#1#2{\relax\ifmmode{}^{+#1}_{-#2}\else$^{+#1}_{-#2}$\fi}
\def\erparen#1#2{\relax\ifmmode{}(^{#1}_{#2})\else$(^{#1}_{#2})$\fi}

\def~{\ifmmode\phantom{0}\else\penalty10000\ \fi}

\def\det{\mathrm{det}}

\def\fm{\mathrm{fm}}
\def\ev{\mathrm{e\kern-0.1em V}}
\def\kev{\mathrm{ke\kern-0.1em V}}
\def\mev{\mathrm{Me\kern-0.1em V}}
\def\gev{\mathrm{Ge\kern-0.1em V}}
\def\tev{\mathrm{Te\kern-0.1em V}}

\def\n#1e#2n{{#1}\times 10^{#2}}

\def\ord#1{\mathcal{O}\left(#1\right)}
\def\nn{\nonumber}

\def\ods2{\mathcal{O}_{\Delta S=2}}
\def\zds2{Z_{\Delta S=2}}

\makeatletter
\def\slash#1{{\mathpalette\c@ncel{#1}}} % TeXbook, bottom of p360
\def\big#1{{\hbox{$\left#1\vbox to1.012\ht\strutbox{}\right.\n@space$}}}
\def\Big#1{{\hbox{$\left#1\vbox to1.369\ht\strutbox{}\right.\n@space$}}}
\def\bigg#1{{\hbox{$\left#1\vbox to1.726\ht\strutbox{}\right.\n@space$}}}
\def\Bigg#1{{\hbox{$\left#1\vbox
to2.083\ht\strutbox{}\right.\n@space$}}}
\makeatother

\makeatletter
\def\old@comma{,}
\catcode`\,=13
\def,{%
  \ifmmode%
    \old@comma\discretionary{}{}{}%
  \else%
    \old@comma%
  \fi%
}
\makeatother

\makeatletter
\def\old@semicolon{;}
\catcode`\;=13
\def;{%
  \ifmmode%
    \old@semicolon\discretionary{}{}{}%
  \else%
    \old@semicolon%
  \fi%
}
\makeatother

\usepackage{xcolor}

\def\qed{\mathrm{QED}}
\def\qedL{\mathrm{QED}_L}
\def\qedTL{\mathrm{QED}_{TL}}
\def\Deltaop#1#2{\left[\Delta^{#1}_{#2}\right]}
\def\R{\mathbb{R}}
\def\T{\mathbb{T}}
\def\Z{\mathbb{Z}}
\def\BZL{\mathrm{BZ}_L}
\def\BZT{\mathrm{BZ}_T}
\def\BZTL{\mathrm{BZ}_{TL}}
\def\TBZL{\frac{2\pi}{L}\Z}
\def\TBZT{\frac{2\pi}{T}\Z}

\DeclareMathOperator*{\sumint}{%
\mathchoice%
  {\ooalign{$\displaystyle\sum$\cr\hidewidth$\displaystyle\int$\hidewidth\cr}}
  {\ooalign{\raisebox{.14\height}{\scalebox{.7}{$\textstyle\sum$}}\cr\hidewidth$\textstyle\int$\hidewidth\cr}}
  {\ooalign{\raisebox{.2\height}{\scalebox{.6}{$\scriptstyle\sum$}}\cr$\scriptstyle\int$\cr}}
  {\ooalign{\raisebox{.2\height}{\scalebox{.6}{$\scriptstyle\sum$}}\cr$\scriptstyle\int$\cr}}
}

\newcommand{\al}{\alpha}
\newcommand{\rh}{\rho}
\newcommand{\si}{\sigma}
\newcommand{\beq}{\begin{equation}}
\newcommand{\eeq}{\end{equation}}
\newcommand{\mr}{\mathrm}

\begin{document}

\vspace*{1cm}
\begin{center}
{\huge\bf Ab initio calculation of\\the neutron-proton mass difference}
\end{center}
\vspace*{2cm}

\noindent
Sz.\ Borsanyi$^{1}$,
S.\ Durr$^{1,2}$,
Z.\ Fodor$^{1,2,3}$,
C.\ Hoelbling$^{1}$, S.\ D.\ Katz$^{3,4}$, S.\ Krieg$^{1,2}$, L.\ Lellouch$^{5}$,
T.\ Lippert$^{1,2}$, A.\ Portelli$^{5,6}$, K.\ K.\ Szabo$^{1,2}$, B.\ C.\ Toth$^{1}$\\
\vspace*{2cm}\\

\noindent
$^{1}$\ Department of Physics, University of Wuppertal, D-42119 Wuppertal, Germany\\
$^{2}$\ J\"ulich Supercomputing Centre, Forschungszentrum J\"ulich, D-52428 J\"ulich, Germany\\
$^{3}$\ Institute for Theoretical Physics, E\"otv\"os University, H-1117 Budapest, Hungary\\
$^{4}$\ Lend\"ulet Lattice Gauge Theory Research Group, Magyar Tudom\'anyos Akad\'emia--E\"otvos Lor\'and University, H-1117 Budapest, Hungary\\
$^{5}$ CNRS, Aix-Marseille Universit\'e, Universit\'e de Toulon, CPT UMR 7332, F-13288, Marseille, France\\
$^{6}$\ School of Physics and Astronomy, University of Southampton, SO17 1BJ, UK\\
\vspace*{3cm}
\\
\\
\noindent The existence and stability of atoms rely on the fact that neutrons are more
massive than protons. The measured mass difference is only 0.14\% of the average of the two masses.
A slightly smaller or larger value would have led to a dramatically
different universe. Here, we show that this difference results from the
competition between electromagnetic and mass isospin breaking effects.
We performed lattice quantum-chromodynamics and quantum-electrodynamics computations with
four nondegenerate Wilson fermion flavors and computed the neutron-proton mass-splitting 
with an accuracy of $300$ kilo-electron volts,
which is greater than $0$ by $5$ standard deviations.
We also determine the splittings in the $\Sigma$, $\Xi$,
$D$ and $\Xi_{cc}$ isospin multiplets, exceeding in some cases the precision of experimental measurements.

\makeatletter
\renewcommand\@biblabel[1]{\it(#1)}
\makeatother
\renewcommand\citeform[1]{{\it #1}}
\renewcommand\citeleft{(}
\renewcommand\citeright{)}

\chapter{}

The mass of the visible universe is
a consequence of the strong interaction~\cite{Henley:2013jga}, which is the force 
that binds together
quarks into protons and neutrons.
To establish this
with percent-level accuracy, very precise calculations based on the lattice 
formulation of quantum chromodynamics(QCD), the theory of
the strong interaction, were needed.
Going beyond such calculations to control much finer effects that are
at the per mil ($\permil$) level is necessary to, for instance, account for the
relative neutron-proton mass difference which was experimentally measured to
be close to 0.14\%~\cite{Beringer:1900zz}. Precisely, this difference
is needed to explain the physical world as we know it today~\cite{Jaffe:2008gd}.
For example, a relative neutron-proton mass
difference smaller than about one third of the observed 0.14\% would cause
hydrogen atoms to undergo inverse beta decay, leaving predominantly neutrons.
A value somewhat larger than 0.05\% would have resulted in the Big Bang 
Nucleosynthesis (BBN),
producing much more helium-4 and far less hydrogen than it did in our
universe. As a result, stars would not have ignited in the way they did.  On
the other hand, a value considerably larger than 0.14\% would have resulted in a much
faster beta decay for neutrons. This would have led to far fewer neutrons at the end
of the BBN epoch and would have made the burning of hydrogen in stars and the
synthesis of heavy elements more difficult. We show here that this tiny mass 
splitting is the result of a subtle
cancellation between electromagnetic and quark mass difference effects.

The Standard Model of Particle Physics is a $SU(3)\times SU(2) \times U(1)$
gauge theory with massless fermions. 
During the expansion of the early universe, the Higgs mechanism 
broke this symmetry down to $SU(3)\times U(1)$ and elementary particles 
acquired masses proportional to their couplings to the Higgs field. 
As the universe continued to expand, a QCD transition took place,
 confining quarks and gluons into hadrons and giving those 
particles most of their mass. This same theory today is believed to be 
responsible for the tiny isospin splittings which are the subject of
this paper. At the level of precision that we aim for here, the effects 
of the weak interaction, of leptons, and of the two heaviest quarks can
either be neglected or absorbed into the remaining parameters of the theory. 
The resulting theory is one of $u$, $d$, $s$ and $c$ (up, down, strange and 
charm) quarks, gluons, photons and their interactions. The Euclidean 
Lagrangian for this theory is ${\cal L} = 1/(4e^2) F_{\mu\nu}F_{\mu\nu}
+1/(2g^2)\text{Tr}G_{\mu\nu}G_{\mu\nu} + \sum_f{\bar \psi_f}[\gamma_\mu 
(\partial_\mu + i q_f A_\mu + i B_\mu) + m_f ] \psi_f$, where $\gamma_\mu$ are the Dirac matrices, 
$f$ runs over 
the four flavors of quarks, the $m_f$ are their masses and the $q_f$ are their 
charges in units of the electron charge $e$. Moreover, 
$F_{\mu\nu}=\partial_\mu A_\nu - \partial_\nu A_\mu$, 
$G_{\mu\nu}=\partial_\mu B_\nu - \partial_\nu B_\mu+[B_\mu,B_\nu]$ and 
$g$ is the QCD coupling constant. In electrodynamics, the
gauge potential $A_\mu$ is the real valued photon field, whereas
in QCD, $B_\mu$ is a Hermitian 3 by 3 matrix field. 
The $\psi_f$ are Dirac-spinor fields representing the quarks and carry a
``color'' index, which runs from 1 to 3.
In the present work, we consider 
all of the degrees of freedom of this Lagrangian; that is, we include quantum electrodynamics (QED) and 
QCD, as well as the four nondegenerate quark flavors, in a fully dynamical 
formulation.

The action $S$ of QCD+QED is defined as the spacetime integral of ${\cal L}$.
Particle propagators are averages of products of fields over all possible field
configurations, weighted by the Boltzmann factor $\exp(-S)$. A notable
feature of QCD is asymptotic freedom, which means that the interaction becomes
weaker and weaker as the relative momentum of the interacting particles
increases~\cite{Gross:1973id, Politzer:1973fx}. 
Thus, at high energies the coupling constant is small, and a perturbative treatment is possible.
However, at energies typical
of quarks and gluons within hadrons, the coupling is large, and the interactions
become highly nonlinear. 
The most systematic way to obtain predictions in this nonperturbative regime of
QCD involves introducing a hypercubic spacetime lattice with lattice
spacing $a$~\cite{Wilson:1974sk} on which the above Lagrangian is
discretized, numerically evaluating the resulting propagators and extrapolating
the results to the continuum ($a\rightarrow0$). The discretization procedure
puts fermionic variables on the lattice sites, whereas gauge fields are
represented by unitary 3 by 3 matrices residing on the links between neighboring sites. The
discretized theory can be viewed as a four-dimensional statistical physics
system.

Calculating the mass differences between the neutral and charged hadron
partners by using lattice techniques has involved different levels of approximation.
In the pioneering work of \cite{Duncan:1996xy}, the
quenched approximation was used both for QCD and QED. Recent
studies~\cite{Blum:2010ym,Basak:2013iw,Borsanyi:2013lga} have
typically performed
dynamical QCD computations with quenched QED fields. Another quenched QED
approach, in which the path integral is expanded to $O(\alpha)$, has also
recently been implemented \cite{deDivitiis:2013xla}. In all such calculations,
the neglected terms are of the same leading order in $\alpha$ as the isospin
splittings of interest~\cite{Borsanyi:2013lga}. To have a calculation that
fully includes QED effects to $O(\alpha)$ requires including electromagnetic
effects in the quark sea. Three exploratory studies have attempted to
include these effects. 
The first two used reweighting techniques in $N_f=2+1$ QCD
simulations~\cite{Aoki:2012st,Ishikawa:2012ix}.
Beyond the
difficulty of  estimating the systematic error associated with reweighting, the
computation in \cite{Aoki:2012st} was carried out with a single lattice spacing
in a relatively small $(3\,\fm)^3$ spatial volume and the one in
\cite{Ishikawa:2012ix} on a single, much coarser and smaller lattice, with pion
masses larger than their physical value. In the third study \cite{Horsley:2013qka},
real dynamical QCD and QED simulations were performed, albeit on a single lattice
at unphysical quark mass values.

Here, we provide a fully
controlled ab initio calculation for these isospin splittings.  We used 1+1+1+1
flavor QCD+QED with 3HEX (QCD) and 1 APE (QED) smeared clover improved
Wilson quarks. Up to now, the most advanced simulations have included up, down,
and strange quarks in the sea but neglected all electromagnetic and up-down
mass difference effects. Such calculations have irreducible systematic
uncertainties of $O(1/N_c/m_c^2,\alpha,m_d-m_u)$, where $N_c=3$ is the number
of colors in QCD. This limits their accuracy
to the percent level. We reduced these uncertainties to
$O(1/N_c/m_b^2,\alpha^2)$, where $m_b$ is the
bottom quark mass, yielding a complete description of the interactions
of quarks at low energy, accurate to the per mil level. 

In our parameter set, we have four lattice spacings ranging from $0.06\,\fm$ to
$0.10\,\fm$. We observed very small cutoff effects in our results, which is in
good agreement with our earlier spectrum determination
\cite{Durr:2008rw,Durr:2008zz}.  Nevertheless, these small cutoff effects are
accounted for in our systematic error analysis as $g^2 a$ or
 $a^2$
 corrections in the histogram method described in~\cite{som}.

We performed computations with four values of the bare fine structure constant:
$0$, a value close to the physical value of $1/137$, and two larger values, approximately
$1/10$ and $1/6$. Most of our runs were carried out at $\alpha=0$ and $\sim\!1/10$.  Because QED effects in typical hadron masses are small (around or below
the $1 \permil$ level), statistical noise in the splittings can be
reduced by interpolating between results obtained with the larger
value of $\alpha$ and those obtained with $\alpha=0$. We then confirmed and
improved this interpolation with simulations near the physical value of the
coupling. The actual interpolation to the physical point is performed in terms
of a renormalized fine structure constant defined via the QED Wilson flow
\cite{Luscher:2010iy}.  Within the precision reached in our work, the
splittings studied show no deviation from linear behavior in the range of
couplings studied. Our largest value of the fine structure constant
was chosen so as to increase the signal for the mass splittings, while keeping
under control large finite-volume corrections of the kind discussed below.

Our smallest pion mass is about $195\;\mathrm{MeV}$ (with more than 20,000
trajectories), and our largest lattice has a spatial extent of $8$~fm. These
parameters were carefully chosen to allow for a determination of the
neutron-proton mass splitting that is $\sim\!5$ standard deviations (SDs) from $0$,
with currently available computing resources.  This is a challenge because
accounting for isospin breaking effects increases the cost \cite{som} of the
calculation compared with computations with two degenerate light flavors used
typically in recent works
\cite{Aoki:2008sm,Ohki:2008ff,Lin:2008pr,Bazavov:2009bb,Durr:2010aw,Baron:2010bv,Bietenholz:2011qq,Arthur:2012opa,Fritzsch:2012wq}.

We produced gauge configurations with an improved version of 
the Hybrid Monte Carlo algorithm and checked, a posteriori, that the 
probability weights are always positive in the region of the 
parameter space used in our simulations.

We used two previously suggested frameworks for the photon fields. These
correspond to a nonlocal modification of the action that vanishes in the
infinite-volume limit. As we argue in \cite{som}, these nonlocalities do not
generate new ultraviolet divergences at one-loop order in $\alpha$. The final
analysis is performed in the framework of \cite{Hayakawa:2008an}, which
respects reflection positivity and has a well-defined, large-time limit, unlike
previously used techniques ~\cite{som}.  Generically, the photon fields show
very large autocorrelation times of several thousand trajectories.  We designed
a Fourier accelerated algorithm within this QCD+QED framework that
dramatically reduces these large autocorrelation times.

The long-range nature of the electromagnetic interaction poses one of the most
serious difficulties of the present work. It induces finite-volume corrections
that only fall off like inverse powers of the linear extent of the system.
These are far more severe than the QCD finite-volume corrections, which are
exponentially suppressed in these dimensions. Exponential corrections can
easily be included in large scale spectrum studies (\cite{Durr:2008zz}). We 
performed an extensive study of the much larger
power-suppressed finite-volume corrections using both one-loop analytical QED
calculations and high-precision QED simulations~\cite{som}. The size and volume 
behavior of these corrections in our full QCD+QED calculation are illustrated in 
Fig.~\ref{fi:qced_fvol}. 

Statistical errors on the mass splittings are calculated by using 2000 bootstrap
samples. The systematic uncertainties on the final results are determined with
our histogram method~\cite{Durr:2008zz}. We considered a wide range of
analyses, each of which provides a valid approach to obtain the physical
splittings from our simulation results, and calculated the associated goodness
of fit. Because these procedures have different numbers of free parameters, we
combined them using the Akaike  information criterion (AIC)~\cite{Akaike:1974} and
obtained a distribution for each splitting. The means of these
distributions are our central values, whereas the widths of the distributions
provide estimates of systematic
uncertainties. This procedure yields conservative errors. 

Our final results for the mass splittings are shown in Fig.
\ref{fi:qced_splittings}. 
A comparison with the results of~\cite{Borsanyi:2013lga} indicates that 
the precision of the signal for $\Delta M_N$ (thus the splitting being non-zero) 
increased from $\sim\!\!1\sigma$ to $5\sigma$.
For the other channels, the improvement is even more pronounced. In addition, the present work
represents a fully-controlled approach, whereas \cite{Borsanyi:2013lga} was based on the 
electroquenched approximation with degenerate light quarks in the sea.
The nucleon, $\Sigma$, and $\Xi$ splittings are consistent
with the Coleman-Glashow relation
$\Delta_\mathrm{CG}\equiv\Delta M_N - \Delta M_\Sigma + \Delta M_\Xi 
=0$ \cite{Coleman:1961jn}. According to our calculation, this relation is fulfilled with an 
accuracy of $130\,\kev$. 
We also computed the individual contributions to the
splittings coming from mass isospin breaking effects ($\alpha=0$, $m_d-m_u\neq
0$) and electromagnetic effects ($m_d-m_u=0$, $\alpha \neq 0$), as defined
in~\cite{som}. The numerical results for all of our results are given in Table
\ref{ta:qced}. 
Because the precision of the experimental result for the nucleon
is far greater than ours, we additionally give the QED and QCD separation
obtained using the experimental value of $M_n-M_p$:
$(M_n-M_p)_\mathrm{QCD}/(M_n-M_p)_\mathrm{QED}=-2.49(23)(29)$.
Last, we used this number in Fig.~\ref{fi:qced_regions} to plot the result of 
the neutron-proton mass splitting
as a function of quark-mass difference and electromagnetic coupling. In
combination with astrophysical and cosmological arguments, this figure can be
used to determine how different values of these parameters would change the
content of the universe. This in turn provides an indication of the
extent to which these constants of nature must be fine-tuned to yield a universe
that resembles ours.

\newpage \vspace*{2cm} \noindent{\bf Acknowledgments:}\\ \\ \\ \noindent
%We thank G.\ Endr\H odi, M.\ Knecht and G.\ Zweig for helpful conversations.
This project was supported by the Deutsche Forschungsgemeinschaft grant SFB/TR55, the Partnership
for Advanced Computing in Europe (PRACE) initiative, the
Gauss Centre for Supercomputing e.V, the European Research Council grant (FP7/2007-2013/ERC No
208740), the Lend\"ulet program of the Hungarian Academy of Sciences (LP2012-44/2012), 
''Origines, Constituants et \'Evolution de l'Univers'' (OCEVU) Labex
(ANR-11-LABX-0060), the A*MIDEX project (ANR-11-IDEX-0001-0) funded by the
''Investissements d'Avenir'' French government program and managed by the Agence Nationale de la
Recherche (ANR), and the Grand \'Equipement National de Calcul Intensif--Institut du 
D\'eveloppement et des Ressources en Informatique Scientifique (IDRIS)
Grand Challenge grant 2012 "StabMat" as well as grant No. 52275.
The computations were performed on JUQUEEN and JUROPA at Forschungszentrum
J\"ulich (FZJ), on Turing at IDRIS in Orsay, on SuperMUC at Leibniz Supercomputing Centre in
M\"unchen, on Hermit at the High Performance Computing Center in Stuttgart and
on local machines in Wuppertal and Budapest. The data described in the paper are 60 TB and archived 
in FZJ.

\renewcommand\thefigure{\arabic{figure}}
\renewcommand\thetable{\arabic{table}}

\newpage
\noindent
\begin{figure}[h!]
\centering
\includegraphics{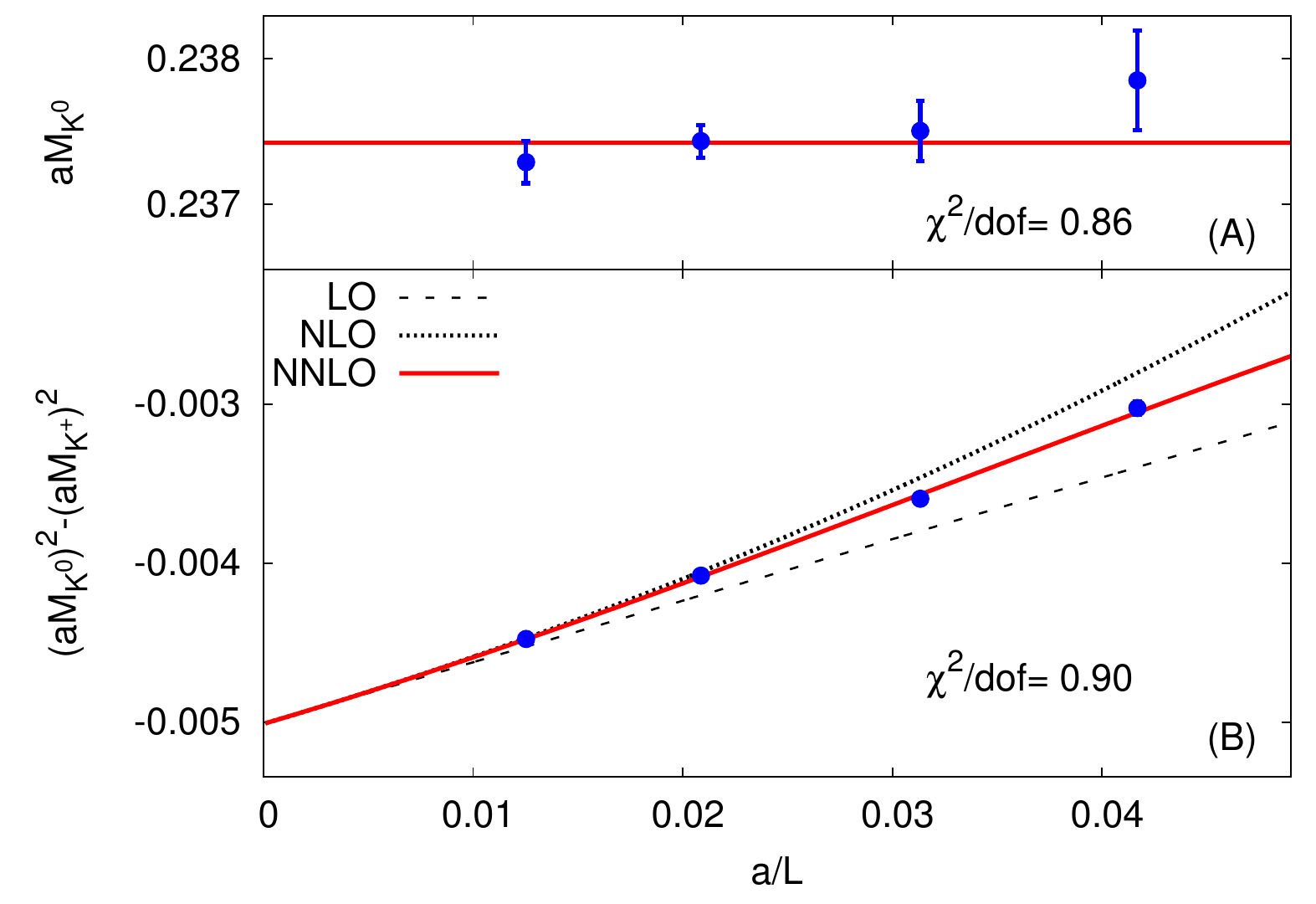}
\caption{\label{fi:qced_fvol} 
{\bf Finite-volume behavior of kaon masses.} ({\bf A})\  The neutral kaon mass, $M_{K^0}$,
shows no significant finite volume dependence; $L$ denotes the linear size of the system. 
({\bf B}) The mass-squared difference of the charged kaon mass,
$M_{K^+}$, and $M_{K^0}$ indicates that $M_{K^+}$ is strongly dependent on volume.
This finite-volume dependence is well described by an asymptotic expansion in $1/L$ 
whose first two terms are 
fixed by QED Ward-Takahashi identities \cite{som}. The solid 
curve depicts a fit of the lattice results (points) to the expansion up to 
and including a fitted $O(1/L^3)$ term. The dashed and dotted curves show the contributions 
of the leading and leading plus next-to-leading order terms, respectively.
The computation was performed by using the following parameters:
bare $\alpha\sim\!1/10$, $M_\pi=290\,\mev$, and $M_{K^0}=450\,\mev$. The mass difference
is negative because a larger-than-physical value of $\alpha$ was used.
The lattice spacing $a$ is $\sim\!0.10\,\fm$.
}
\end{figure}

\newpage
\noindent
\begin{figure}[h!]
\centering
\includegraphics*[width=16cm]{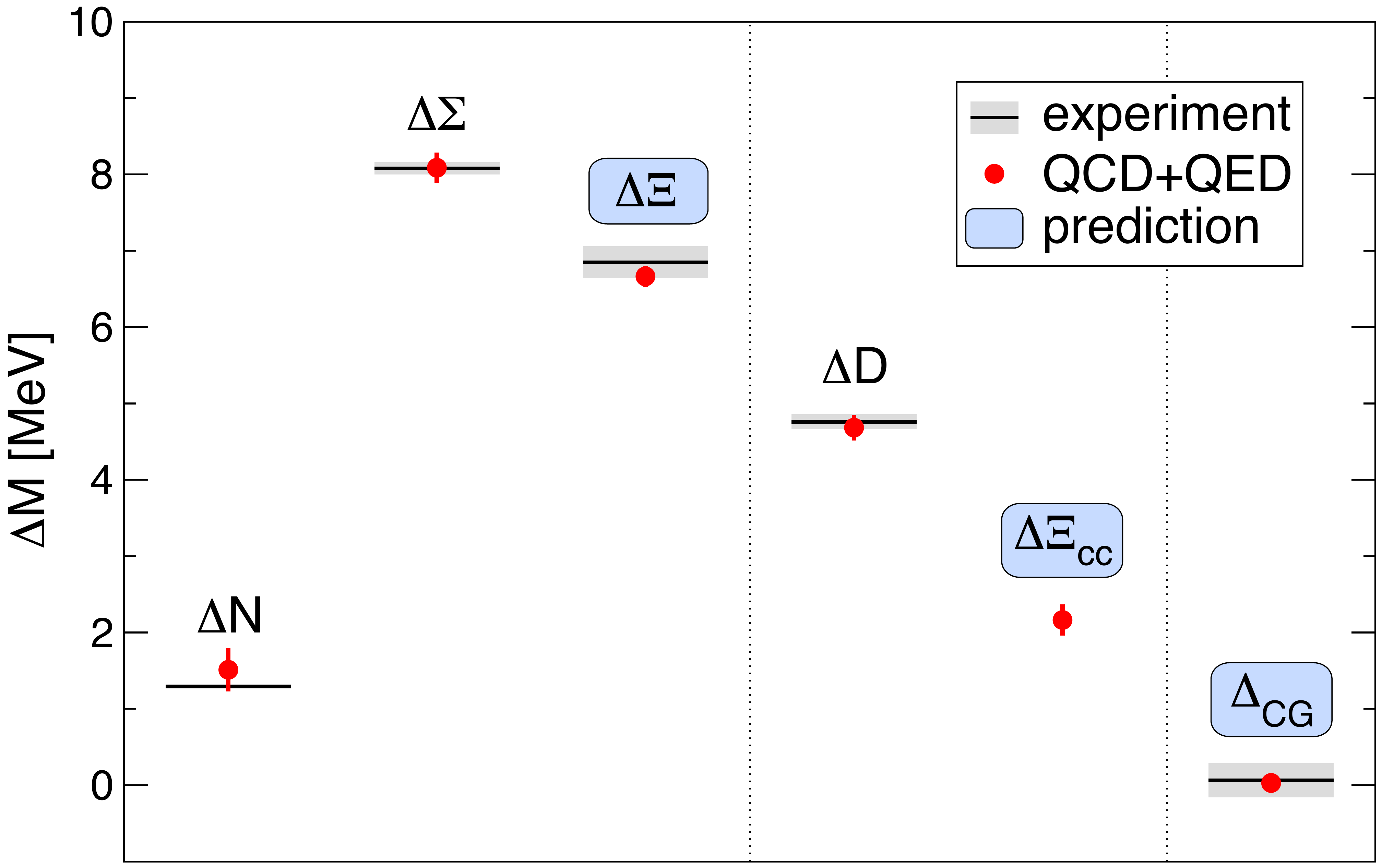}
\caption{\label{fi:qced_splittings}
{\bf Mass splittings in channels that are stable under the strong and
electromagnetic interactions.} Both of these interactions are fully unquenched in
our 1+1+1+1 flavor calculation. The horizontal lines are the experimental values
and the grey shaded regions represent the experimental error \cite{Beringer:1900zz}.
Our results are shown by red dots with their uncertainties.
The error bars are the squared sums of the
statistical and systematic errors. 
The results for the $\Delta M_N$,
$\Delta M_\Sigma$, and $\Delta M_D$ mass splittings are
post-dictions, in the sense that their values are known experimentally with
higher precision than from our calculation. On the other hand, our calculations
yield $\Delta M_\Xi$, $\Delta M_{\Xi_{cc}}$ splittings, and the Coleman-Glashow 
difference $\Delta_\mathrm{CG}$,
which have either not been measured in experiment or are measured
with less precision than obtained here. This feature is represented by a blue shaded region around the label.
}
\end{figure}

\newpage
\noindent
%\vspace*{1.5cm}\\
\begin{figure}[h!]
\centering
\includegraphics*[width=16cm]{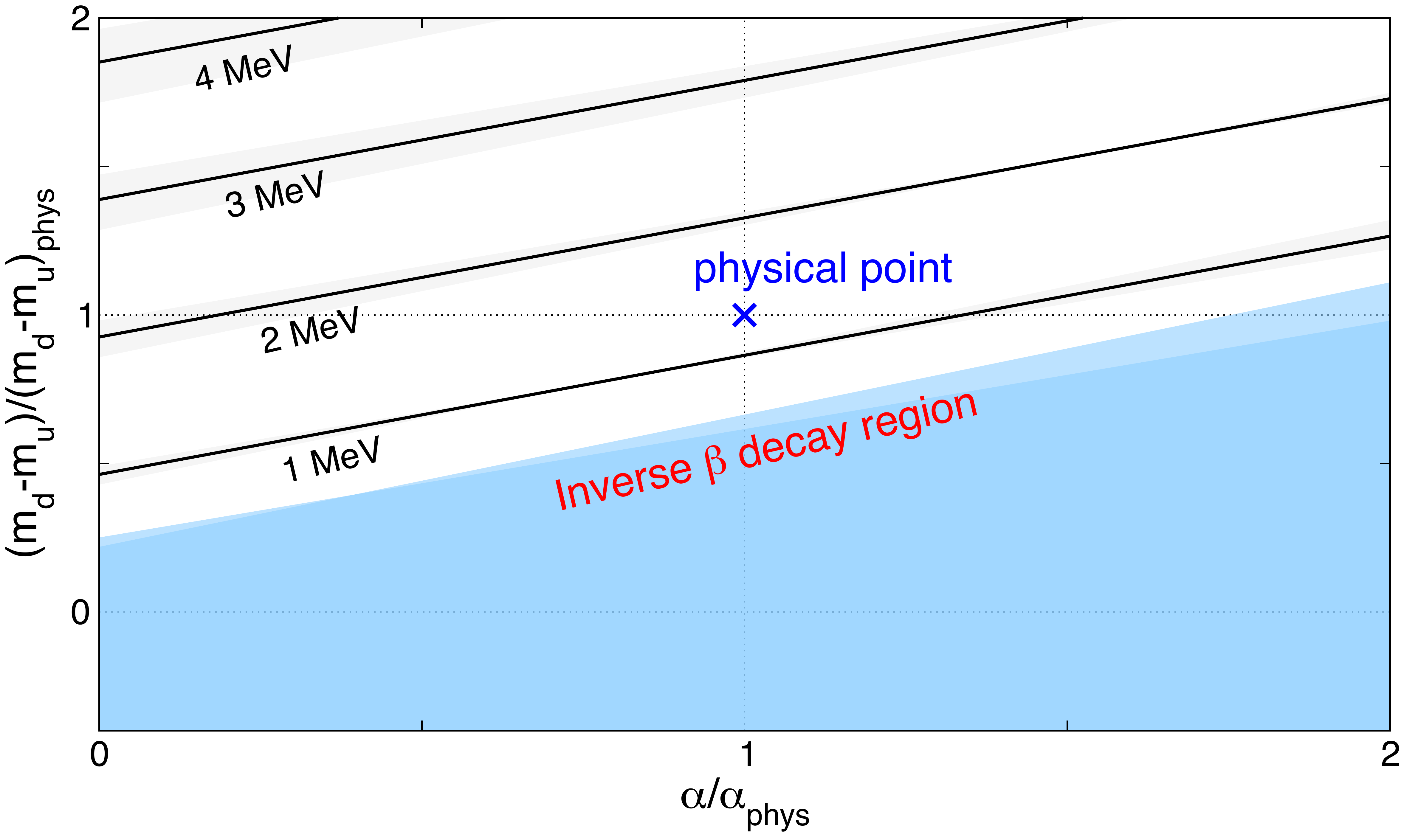}
\caption{\label{fi:qced_regions} 
{\bf Contour lines for the neutron-proton mass 
splitting.} The contours are shown as a function of the quark mass difference
and the fine structure constant, both normalized with their real world, physical 
value. Because these two effects 
compete, by increasing $\alpha$ at fixed quark mass difference one can
decrease the mass difference between the neutron and the proton to $0.511\;\mathrm{MeV}$,
at which inverse $\beta$-decay sets in, as depicted by the blue region.
The blue cross shows the physical point. 
The shaded bands around the contours 
represent the total statistical and systematic uncertainties on these
predictions.
A constraint on
the neutron-proton mass difference obtained from other considerations
leads to a constraint on $m_d-m_u$ and/or $\alpha$, which can be directly read off from the figure.
} 
\end{figure}

\newpage
\noindent
\begin{table}[h!]
\begin{center}
\begin{tabular}{|c|c|c|c|}
\hline
& mass splitting [MeV]& QCD [MeV] & QED [MeV] \\
\hline
$\Delta N= n-p$  & 1.51(16)(23) &  2.52(17)(24) & -1.00(07)(14) \\
$\Delta \Sigma=\Sigma^- - \Sigma^+$ &  8.09(16)(11) &  8.09(16)(11) &0\\
$\Delta \Xi= \Xi^- - \Xi^0$  &   6.66(11)(09) &  5.53(17)(17) &  1.14(16)(09) \\
\hline
$\Delta D= D^{\pm} - D^0$        & 4.68(10)(13) &  2.54(08)(10) &  2.14(11)(07) \\
$\Delta \Xi_{cc} = \Xi_{cc}^{++}-\Xi_{cc}^{+}$ &  2.16(11)(17) &  -2.53(11)(06) & 4.69(10)(17) \\
\hline
$\Delta_\mathrm{CG}=\Delta N-\Delta \Sigma+\Delta \Xi$& 0.00(11)(06) & -0.00(13)(05) &  0.00(06)(02)\\
\hline
\end{tabular}
\end{center}
\caption{\label{ta:qced} 
{\bf Isospin
 mass splittings of light and charm hadrons.} Also shown are the
individual contributions to these splittings from the mass difference
$(m_d-m_u)$ (QCD) and from electromagnetism (QED). The
separation requires fixing a convention, which is described in \cite{som}.
The last line is the violation of the Coleman-Glashow relation \cite{Coleman:1961jn}, which
is the most accurate of our predictions.
}
\end{table}

\newpage
\vspace*{1cm}
\begin{center}
{
    \LARGE\bf
    Supplementary Materials for\\
    \ \\
    Ab initio calculation of the neutron-proton mass difference\\
}
\end{center}
\vspace*{2cm}

\noindent Sz.\ Borsanyi$^{1}$,
S.\ Durr$^{1,2}$,
Z.\ Fodor$^{1,2,3}$,
C.\ Hoelbling$^{1}$, S.\ D.\ Katz$^{3,4}$, S.\ Krieg$^{1,2}$, L.\ Lellouch$^{5}$,
T.\ Lippert$^{1,2}$, A.\ Portelli$^{5,6}$, K.\ K.\ Szabo$^{1,2}$, B.\ C.\ Toth$^{1}$\\
\\
$^{1}$\ Department of Physics, University of Wuppertal, D-42119 Wuppertal, Germany\\
\\
$^{2}$\ J\"ulich Supercomputing Centre, Forschungszentrum J\"ulich, D-52428 J\"ulich, Germany\\
\\
$^{3}$\ Institute for Theoretical Physics, E\"otv\"os University, H-1117 Budapest, Hungary\\
\\
$^{4}$\ MTA-ELTE Lend\"ulet Lattice Gauge Theory Research Group, H-1117 Budapest, Hungary\\
\\
$^{5}$ CNRS, Aix-Marseille Universit\'e, Universit\'e de Toulon, CPT UMR 7332, F-13288, Marseille, France\\
\\
$^{6}$\ School of Physics \& Astronomy, University of Southampton, SO17 1BJ, UK\\
\begin{center}
Correspondence to: fodor@physik.uni-wuppertal.de
\end{center}
\vspace*{1cm}

\noindent{\bf This PDF File includes:}\\
\hspace*{1cm} Methods in Sections 1 to 12\\
\hspace*{1cm} Figures S1 to S12\\
\hspace*{1cm} Tables  S1 to S5\\

\newpage
\setcounter{figure}{0}
\setcounter{table}{0}
\renewcommand\thefigure{S\arabic{figure}}
\renewcommand\thetable{S\arabic{table}}
\renewcommand{\theequation}{S\arabic{equation}}
\makeatletter
\renewcommand\@biblabel[1]{\it(#1)}
\makeatother
\renewcommand\citeform[1]{{\it #1}}
\renewcommand\citeleft{(}
\renewcommand\citeright{)}

\makeatletter
\@fpsep\textheight
\makeatother

\section{Outline}

In the following sections we provide details of the work presented in the
main paper.  In \sec{se:lattice} we define the theory that we use, namely QCD
and QED with four quark flavors on a four dimensional lattice. After fixing
our notations we discuss the action for the photon field in \sec{se:phact}.
There are many subtleties here, such as gauge fixing and zero-mode
subtraction. In \sec{se:dirac} we define our Dirac operator, which contains
both photon and gluon fields. We apply one step of APE smearing to the
electromagnetic field and 3 steps of HEX smearing to the $SU(3)$ field. In
\sec{se:hex} we discuss in detail what the advantages of these choices are
and how we optimized the smearing parameters.

The determination of the mass splittings for the isospin multiplets needs a
careful treatment of QED on the lattice.  In \sec{se:qedana} we discuss in
detail the differences between two possible formulations of lattice QED
\cite{Hayakawa:2008an} and illustrate numerically the disadvantages of
the one used in all previous numerical studies in \sec{se:qednum}. We also
compare our numerical implementation of lattice QED with lattice
perturbation theory up to $\mathcal{O}(e^2)$ order \cite{Mertens:1997wx} and
recover the tiny $\mathcal{O}(e^4)$ corrections.

The photon fields are long-ranged and the mass corrections are
proportional to $1/L$, $1/L^2$,
\dots\ types, where $L$ is the spatial size of the system. These are much
larger corrections than those in QCD, which are exponentially suppressed for
stable particles. Actually they are of the same
order as the mass-splittings themselves. 
In \sec{se:qedana} we determine
the finite-volume corrections analytically 
for point particles.
In \sec{se:fvol} the study is generalized to the case of
composite particles. Using the Ward-Takahashi identities we show that
the coefficients of the $1/L$ and $1/L^2$ terms are
universal~\footnote{While we were writing up the results of the
      present paper, an analytical calculation of finite-volume effects in
      a non-relativistic effective field theory framework was presented
      \cite{Davoudi:2014qua}. We agree on the universality of the $1/L$ and $1/L^2$
      coefficients and on their values. However, simplifying the
      composite-particle results of \cite{Davoudi:2014qua} to the point-particle case leads
      to a coefficient of the $1/L^3$ term which differs from the one that
      we obtain in \sec{se:qedana}. We confirm the validity of the latter with
  high-precision QED simulations in that section. } and we
compare these analytical findings to our numerical QCD+QED simulation
results.

Starting with \sec{se:alg} we present the details of the many simulations that
are performed and summarized here. The use of Rational Hybrid Monte-Carlo
method is discussed with a special emphasis on the lowest eigenvalues of the
Dirac operator. Autocorrelations are under control for our choice of parameters
in the QCD part of our work.  However, due to the zero mass of the photon and
the correspondingly large correlation lengths, a standard Hybrid Monte-Carlo
integration of the photon fields results in large autocorrelation times. We
show how we solved this problem by developing a Fourier accelerated algorithm.
For the propagator calculations we used a 2-level multi-grid approach to have
several hundred source positions and significantly improve our statistics.  We
present the ensembles generated for this project in \sec{se:ens}. We use four
different lattice spacings in the range of $0.06$ to $0.10\,\fm$,
and pion masses down to $195\,\mev$. We have runs
with zero electromagnetic coupling and with non-zero ones. Altogether we have
accumulated 41 ensembles, whose parameters are detailed in that section.  In
\sec{se:qflow} we present our renormalization prescription for the electric
charge.

In the final sections, \secss{se:iso}{se:aic}{se:fit}, we detail the procedure
that is used to extract mass splittings. We explain how we separate the QED and
quark-mass-difference contributions to these differences. In obtaining our
final results, we conduct a thorough investigation of systematic
uncertainties. To determine these, we use the histogram method
\cite{Durr:2008zz}. We extend it by using the Akaike's information criterion
(AIC).

\section{Lattice and action details}
\label{se:lattice}

The elementary particles that we consider in this paper are photons
($A$), gluons ($U$), and the up, down, strange and charm quarks
($\psi_f$ with $f={u,d,s,c}$). The contributions from other known
elementary particles to the isospin splittings can be neglected given
the accuracy required for our study.
The following action describes the interactions the degrees of freedom kept here:
\[
S[U,A,\overline{\psi},\psi]=
S_g[U;g] + S_\gamma[A] +
\sum_{f} \overline{\psi}_f D[U,A;e,q_f,m_f]\psi_f.
\]
where $S_g$ and $S_\gamma$ are the gluon and photon actions, respectively, and $D$ is the
Dirac operator. This action has the following parameters: 
the gluon gauge coupling $g$,
the electromagnetic coupling $e$, the four quark masses $m_f$ and the four 
charge parameters $q_u=q_c=2/3$ and $q_d=q_s=-1/3$.

We work on a four dimensional, cubic, Euclidean lattice with $L$ points in the spatial
and $T$ points in the time direction.
The boundary condition is periodic for the photon and gluon fields.
For the quark fields it is periodic in the spatial directions and anti-periodic
in time. Lattice fields in coordinate space $f_x$ can be
transformed into momentum space $f_k$ and back by Fourier transformation.
To avoid cumbersome notations we use the same symbol for the Fourier-transformed fields.
For periodic fields \[
k_\mu=
\begin{cases}
2\pi/aT\cdot\{0,1,\dots,T-1\} & \mu=0,\\
2\pi/aL\cdot\{0,1,\dots,L-1\} & \mu=1,2,3.
\end{cases} \]
\noindent The discrete differential of a lattice field $f_x$ is defined as $\partial_\mu f_x= (f_{x+a\mu}-f_x)/a$,
and its adjoint is given by $\partial_\mu^\dag f_x= (f_x-f_{x-a\mu})/a$. These
differentials are given by multiplications in momentum space:
$\hat{k}_\mu f_k$ and $-\hat{k}^*_\mu f_k$, where the components of the complex $\hat{k}$ vector are defined as
\be
\hat{k}_\mu= \frac{\exp(iak_\mu)-1}{a}.
\label{eq:khatmu}
\ee
For the discretized gluon action $S_g$ we choose the tree-level improved Symanzik action \cite{Luscher:1984xn}. Its
properties are well known and will not be discussed here. 

In order to eliminate discretization artefacts, we carry out a continuum
limit. In QED this is a subtle issue, although without practical relevance
here.  When we talk about continuum limit in this work, we always mean a
limiting procedure where the lattice spacing does not go exactly to zero, but
to a minimal value. This value can be chosen to be extremely small for the
coupling values considered in this work, and the remaining lattice artefacts
are completely negligible. The uncertainties associated with extrapolating
results to this minimal value are orders of magnitude larger and are duly
accounted for in our analysis.

\subsection{Photon action}
\label{se:phact}

This section gives the derivation of the photon action $S_\gamma$. We use a non-compact formulation.
The naive photon action 
\[
S_\gamma^{naive}[A]= \frac{a^4}{4} \sum_{\mu,\nu,x} (\partial_\mu A_{\nu,x}-\partial_\nu A_{\mu,x})^2
\]
is left invariant by gauge transformations with a field $f_x$:
\[
A_{\mu,x} \to A_{\mu,x} - \partial_\mu f_x.
\]
Field modes that are generated by these transformations,
appear neither in the naive photon action nor 
in the part of the action which
describes the coupling to the quarks. As a result, in
the non-compact formulation gauge-variant observables, such as charged 
particle propagators, are ill-defined.
To avoid this, one chooses a gauge (this would also be needed in the compact formulation). We use the Coulomb gauge for the
photon field in this work. 

There is another set of symmetry transformations of the naive photon action
that
shift the photon field by a constant $c_\mu$:
\[
A_{\mu,x} \to A_{\mu,x} + c_\mu.
\]
Because of our use of periodic boundary conditions, this
symmetry is not a gauge symmetry. However,  in the infinite-volume limit it
becomes a gauge symmetry with $f_x= -c_\mu x_\mu$. The treatment of
this symmetry requires special attention, which we detail presently.

\subsubsection*{Zero mode subtraction}

We eliminate the shift symmetry of the naive photon action by removing the
zero-momentum mode of the photon field from the path integral. This step is not
strictly necessary, since only a discrete subset of the shift transformation is a symmetry of the
quark action. However significant complications were observed in simulations
with a non-vanishing zero-mode \cite{Gockeler:1991bu}. Additionally the zero-mode removal makes the theory well-defined perturbatively. The removal of
modes, that form a set of measure zero in the infinite-volume limit, is a
legitimate procedure, since it does not effect the path integral in this limit.
There are different proposals in the literature for removing the zero mode. These
correspond to different realizations of the theory in finite volume.

The simplest procedure is to set
\begin{align}
a^4\sum_x A_{\mu,x}=0 \text{  for all
$\mu$.}
\label{eq:zm1}
\end{align}
The sum runs for the temporal and spatial directions, so we will denote this choice $\qedTL$.
This setup is the one used in all previous studies which include QED corrections
to hadronic observables in lattice QCD
\cite{Duncan:1996xy,Blum:2007cy,deDivitiis:2013xla,Borsanyi:2013lga}. The
disadvantage of this choice is, that it violates reflection positivity. This
can be seen by adding the zero-mode constraints to the path integral in the
following form: 
\[\lim_{\xi\to0} \exp \left[-\sum_\mu (a^4\sum_x
A_{\mu,x})^2/\xi^2\right].
\] 
A $\left(a^4\sum_x A_{\mu,x}\right)^2$ term in the
action, which connects fields at arbitrary positive and negative times, spoils
reflection positivity. It has a serious consequence: charged particle
propagators are ill-behaved, if the time extent of the box is sent to infinity while keeping its spatial size fixed.
This is demonstrated both analytically and numerically in \secs{se:qedana}{se:qednum}. 

Another choice, proposed by Hayakawa and Uno \cite{Hayakawa:2008an}, is to remove
the zero mode of the field on each time slice separately:
\begin{align}
a^3\sum_{\vec{x}} A_{\mu,x_0,\vec{x}}=0 \text{  for all
$\mu$ and $x_0$.}
\label{eq:zm2}
\end{align}
The sums run here only over the spatial directions and this prescription will
be denoted $\qedL$.  This constraint can be shown to be reflection positive and
charged-particle propagators are well behaved in this case. Thus particle
masses can be extracted from the large time behavior of these propagators.  We
study both prescriptions in detail in this paper and compare them.  The main
results of the paper are obtained in the $\qedL$ formulation.

Although $\qedL$ represents a nonlocal modification of the path-integral and
violates hypercubic symmetry, these effects vanish in physical quantities
in the infinite volume limit. In finite volume an important issue is the
renormalizability of the theory.  We will show in \sec{se:qedana} for the
case of point-like particles and argue in \sec{se:fvol} for the case of composite ones, that the
divergences in the one-loop self-energy are the
same in $\qedL$ and in the infinite volume theory. No new counterterms are
required for this particular diagram, which is the relevant one, when determining
electromagnetic corrections to the masses. We expect, that this property holds
for other quantities as well, although this has to be checked by explicit calculations, similar
to the ones, that are presented in this paper.

\subsubsection*{Coulomb gauge via Feynman gauge}

In order to ensure the existence of a transfer matrix it is convenient and
usual to choose the Coulomb gauge.
After removing the zero mode on each time slice, the Coulomb-gauge fixing
condition $\vec{\nabla}^\dag\cdot \vec{A}_x=0$ defines a unique operator $P_C$, that
transforms a field configuration into Coulomb gauge. The transformation
in momentum space is given by:
\begin{align}
A\to A'=P_CA
\quad
\text{with}
\quad
P_{C,\mu\nu}=
\delta_{\mu\nu}-|\vec{\hat{k}}|^{-2} \hat{k}_\mu(0,\vec{\hat{k}}^*)_\nu,
\label{eq:pc}
\end{align}
with $\hat{k}_\mu$ given in \eq{eq:khatmu}. Generating field configurations in Coulomb gauge in the full dynamical case
would be somewhat cumbersome. We therefore decided to generate configurations
in Feynman gauge using the action
\begin{align}
\label{eq:sgamma}
S_\gamma[A]= \frac{1}{2TL^3}\sum_{\mu,k} |\hat{k}|^2 |A_{\mu,k}|^2
\end{align}
and then transform them into Coulomb gauge using the $P_C$ operator. It can be proven
that this is equivalent to using Coulomb gauge directly.

\subsection{The Dirac operator}
\label{se:dirac}

The Wilson operator with tree-level clover improvement~\cite{Sheikholeslami:1985ij} is chosen as our
lattice Dirac operator. The starting point is the
gauge-covariant Dirac operator, which acts on a spinor field, $\psi$, as follows:
\begin{gather*}
\left(D[U,A;e,q,m]\psi\right)_x= \\
(\frac{4}{a}+m)\psi_x -\\
-\frac{1}{2a}\sum_\mu \left[ (1+\gamma_\mu)\exp(ieqa\tilde{A}_{\mu,x})\tilde{U}_{\mu,x}\psi_{x+\mu} +
(1-\gamma_\mu)\exp(-ieqa\tilde{A}_{\mu,x-\mu})\tilde{U}^\dagger_{\mu,x-\mu}\psi_{x-\mu}\right] +\\
-\frac{i a}{4} \sum_{\nu>\mu} \left(F^{(\tilde{U})}_{\mu\nu,x} + eqF^{(\tilde{A})}_{\mu\nu,x}\right)[\gamma_\mu,\gamma_\nu] \psi_x.
\end{gather*}
The gauge-invariance of the quark action is ensured by exponentializing the
non-compact photon fields. 
We use the MILC convention for the gamma matrices $\gamma_\mu$ \cite{MILC}. Note also that
this Dirac operator differs from the usual definition by an extra minus sign
in front of the $\gamma$'s. $F_{\mu\nu,x}^{(U)}$ is the usual
discretization of the gluon field strength tensor and
is built up from the products of the gluon links along the ``clover'' path. 
$F_{\mu\nu,x}^{(A)}$ is a discretization of the electromagnetic field strength
tensor. It is chosen as the sum of the photon fields around a two-by-two
plaquette centered at $x$ in the $\mu-\nu$ plane.  Gluon and photon fields
($\tilde{U}$ and $\tilde{A}$) that enter the Dirac operator are obtained
by smearing the original gluon and photon fields.  In this work the gluon
fields have undergone three levels of HEX smearing. The parameters of the
HEX smearing procedure are chosen with care, as described in \sec{se:hex}. 
We smear the photon fields with the following transformation
\[
A_{\mu,x}\to \tilde{A}_{\mu,x}= 0.9 \cdot A_{\mu,x}
+ 0.1 \cdot \sum_{\pm\nu\ne\mu} \left( A_{\nu,x}+A_{\mu,x+\nu}-A_{\nu,x+\mu}\right),
\]
where $A_{-\nu,x}=-A_{\nu,x-\nu}$. 

The most important advantage of smearing 
is the reduction of the additive quark mass renormalization. In our case, 
it has two contributions: 
one stems from non-trivial gluon fields, the other
is due to the presence of photons. Although a large additive renormalization is not a problem of
principle, a small one facilitates tuning the parameters in
dynamical simulations.  For illustration of the effect we define the
electromagnetic mass renormalization $\delta$ by a neutral mesonic state, which
is obtained by dropping the disconnected part of the propagator of the
quark--anti-quark system. $\delta$ is defined as the shift in the bare
quark mass to get the same meson mass as in the $e=0$ case, see also Ref.\ \cite{Portelli:2010yn}.  At a lattice
spacing of $0.10\,\fm$ and a coupling of $e=1$ the shift for the up quark is about
$\delta_u=-0.070$ without photon smearing, which is about the same size as the
additive mass renormalization coming from the gluons. With our smearing recipe
we have a four times smaller value $\delta_u=-0.017$.

As any smearing, our choices for the photon and gluon fields change the
results by effects that disappear in the continuum limit. In
\secs{se:hex}{se:qednum} we demonstrate the advantages of our smearing choices.

\subsection{HEX smearing}
\label{se:hex}

In our study, the gluon fields $U$ have undergone three levels of HEX smearing
\cite{Capitani:2006ni}.  The smearing procedure replaces the original gluon
fields $U$ with the HEX smeared $\tilde{U}$:
\bea
V_{\mu,\nu\rh,x}\!&\!=\!&\!\exp
\Big(\frac{\rh_1}{2}\sum_{\pm\si\neq\mu,\nu,\rh}\Big\{\big[
U_{\sigma,x}\,
U_{\mu,x\!+\!\si}\,
U_{\sigma,x\!+\!\mu}^\dag\,
U_{\mu,x}^\dag-\mr{h.c.}\,\big]-\frac{1}{3}\mr{Tr}[.]\Big\}
\Big)U_{\mu,x},
\nonumber\\
W_{\mu,\nu,x}\!&\!=\!&\!\exp
\Big(\frac{\rh_2}{2}\sum_{\pm\rh\neq\mu,\nu}\Big\{\big[
V_{\rh,\mu\nu,x}\,
V_{\mu,\nu\rh,x\!+\!\rh}\,
V_{\rh,\mu\nu,x\!+\!\mu}^\dag\,
U_{\mu,x}^\dag-\mr{h.c.}\,\big]-\frac{1}{3}\mr{Tr}[.]\Big\}
\Big)U_{\mu,x},
\nonumber\\
\tilde{U}_{\mu,x}\!&\!=\!&\!\exp
\Big(\frac{\rh_3}{2}\sum_{\pm\nu\neq\mu}\Big\{\big[
W_{\nu,\mu,x}\,
W_{\mu,\nu,x\!+\!\nu}\,
W_{\nu,\mu,x\!+\!\mu}^\dag\,
U_{\mu,x}^\dag-\mr{h.c.}\,\big]-\frac{1}{3}\mr{Tr}[.]\Big\}
\Big)U_{\mu,x}.
\label{eq:def_HEX}
\eea
For further details on our implementation see~\cite{Durr:2010aw}.

\begin{figure}[H]
\centering
\includegraphics{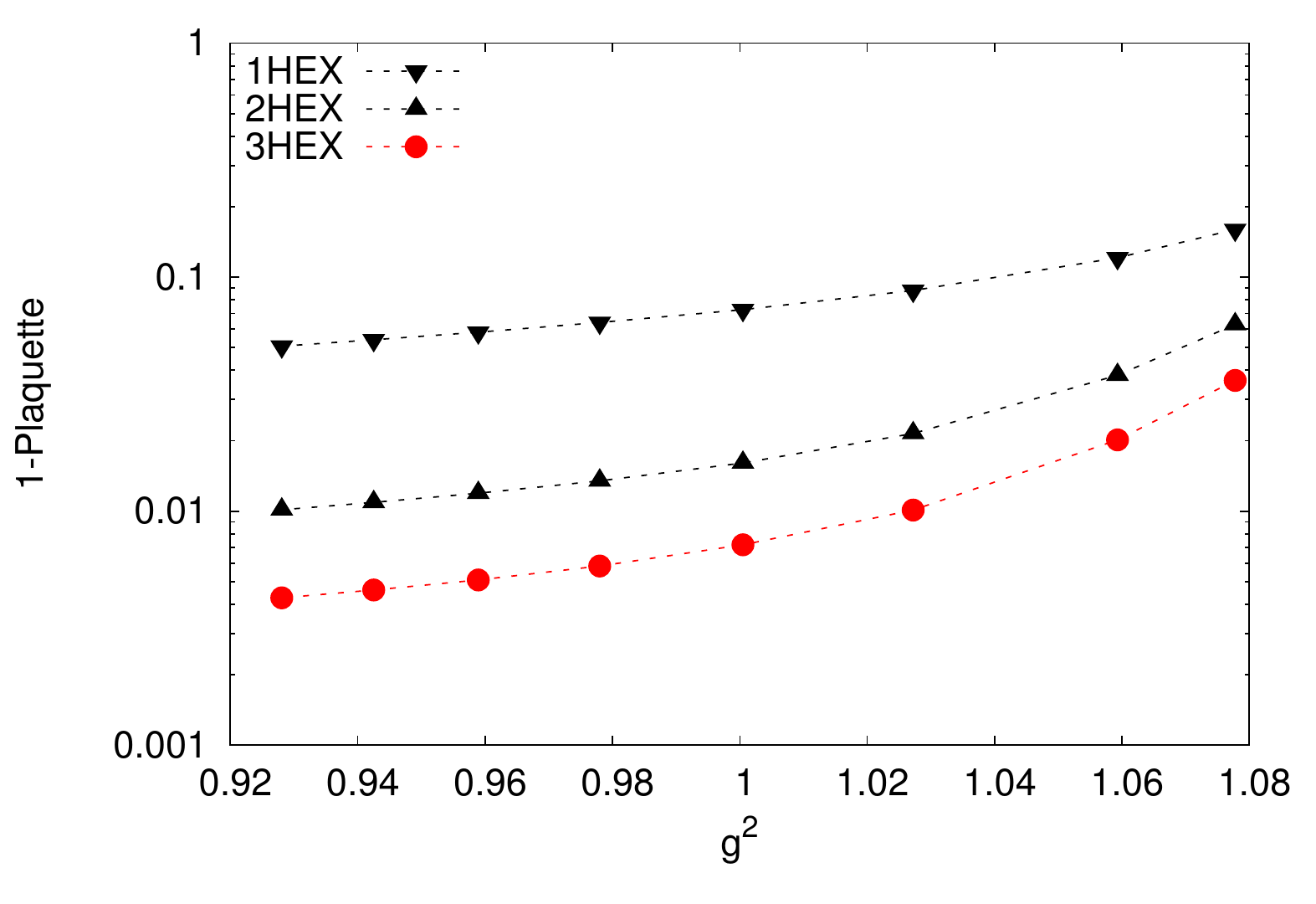}
\caption{\sl\label{fi:hex}
$1-\langle U_\square\rangle$ for 1, 2, and 3 HEX smearing steps,
using our preferred values for the parameters $[\rh_1,\rh_2,\rh_3] = (0.22,0.15,0.12)$. The lines connecting
the points are to guide the eye.
}
\end{figure}

Here, we illustrate our procedure of iterated HEX smearings on lattices
generated in the pure-gauge case with a Wilson action.  We generated sets of matched
lattices with a fixed box length in units of the Sommer scale $L/r_0=3$. We
used the formula for $r_0/a$ as a function of the gauge coupling from 
\cite{Durr:2006ky}, which is based on data from \cite{Necco:2001xg}.  A
wide range of lattice spacings were covered from $a=0.245$ fm down to $a=0.046$
fm.  We consider the three-fold HEX smeared plaquette as a function of $g^2$ on
these lattices with special attention to the $g^2\to0$ behavior.
Since the last step in \eq{eq:def_HEX} is standard
stout smearing, we restrict ourselves to $\rh_3=0.12$. This value has been
used in many studies based on stout smearing in the past and is considered safe
by perturbative considerations \cite{Capitani:2006ni}.  Hence, only two
parameters need to be tuned to optimize the scaling to the continuum limit. 
We selected the value $\rh_\mr{HEX}=(0.22,0.15,0.12)$ as our preferred
HEX-smearing parameter. These parameters correspond to
$\al_\mr{HYP}=(0.44,0.60,0.72)$ in the HYP-smearing scheme.
In \fig{fi:hex} we show results for the average plaquette, $\langle
U_\square\rangle$, for three different smearing levels.
$\langle U_\square\rangle$ approaches $1$ monotonically in the continuum limit.

\section{Analytical studies of various QED formulations in finite volume}
\label{se:qedana}
In this section we derive a number of important results concerning the
properties of the pole mass of a charged particle in various
formulations of QED on a finite spacetime volume. We focus here on point
particles, because the main features are already present in this
simpler situation. For the formulation of QED that we use in our
simulations, in \sec{se:fvol} we investigate the modifications to
these calculations which result from the fact that mesons and baryons
have internal structure.

It is important to have a solid analytical handle on QED finite-volume
(FV) corrections, because they are expected to be large due to the
long-range nature of the electromagnetic interaction. Unlike QCD, QED
has no gap and the photon remains massless even in the presence of
interactions. While the gap in QCD guarantees that FV corrections fall
off exponentially in $LM_\pi$ for sufficiently large
$LM_\pi$ \cite{Luscher:1985dn}, in the presence of QED, quantities are
much more sensitive to the volume and topology of spacetime. It is the
main characteristics of this sensitivity which concerns us in this
section. We use the computed analytical expressions in two important
ways. The first is to decide on the finite-volume formulation of QED
to use in our numerical work. The second is to test our implementation
of QED and the corresponding codes.

The work presented in this paper is concerned with spin-$1/2$ baryons
and spin-$0$ pseudoscalar mesons. Thus we compute the FV corrections
in spinor and scalar QED. Our photon field has periodic boundary
conditions, while the quark fields are periodic in space and
antiperiodic in time. Therefore, baryon fields are antiperiodic in
time and periodic in space, while meson fields are periodic in all
directions. As a result, the topology of our spacetime is the
four-torus, $\T^4$, up to a twist for baryons in the time
direction. Note that for corrections in inverse powers of the torus
size, only the photon boundary conditions are relevant.

As discussed in \sec{se:phact}, we consider two different versions of FV QED:
\begin{itemize}

\item the first where only the four-momentum zero-mode of the
  photon field is eliminated, i.e.\ $A_\mu(k=0)\equiv 0$, which
  we denote $\qedTL$;

\item the second where all three-momentum zero-modes of the photon
  field are eliminated, i.e.\ $A_\mu(k_0,\vec{k}=\vec 0)\equiv
  0$ for all $k_0$, which we denote $\qedL$.

\end{itemize}

Power-like FV corrections arise from the exchange of a photon around
the torus. They are obtained by comparing results obtained in FV with
those of our target theory, QED in infinite volume (IV), that is in
$\R^4$. Here we are interested in the FV corrections to a charged
particle's pole mass. This is the physical mass of the particle, as
obtained by studying the Euclidean time-dependence of a relevant, zero
three-momentum, two-point correlation function. This mass is gauge
invariant and we use this freedom to work in the simpler Feynman
gauge.

The FV corrections to the mass $m$ of a point particle of spin $J$ and of
charge $q$ in units of $e$, on a torus of dimensions $T\times L^3$, is
given by the difference of the FV self energy,
$\Sigma_J(p,T,L)$, and its IV counterpart, $\Sigma_J(p)$, on shell:
\bea
\Delta m^{n_J}_J(T,L)\equiv m^{n_J}_J(T,L)-m^{n_J} &=& (qe)^2\Delta\Sigma_J(p=im,T,L)\nn\\
& \equiv & (qe)^2\left[\Sigma_J(p=im,T,L) - \Sigma_J(p=im)\right]
\ ,
\labell{eq:Dmn}
\eea
where $n_J=1$ (resp.\ $n_J=2$) for spin $J=1/2$ fermions (resp.\ spin
$J=0$ bosons) and $p=im$ is a shorthand for $p=(im,\vec{0})$ (with
$\slashed{p}\to im$ for spin-$1/2$ fermions). Here and below,
quantities without the arguments $L$ and $T$ are infinite
spacetime-volume quantities.

Because we only work in a regime where electromagnetic effects
are linear in the fine structure constant $\alpha$, we evaluate the
self-energy difference in \eq{eq:Dmn} at one loop. At this order, we
generically write differences of self energies or of contributions to
self energies as
\be
\Delta\Sigma(p,T,L)
=
\left[\sumint_k^\prime-\int\frac{d^4k}{(2\pi)^4}\right]
\sigma(k,p)
\ ,
\labell{eq:DSig1loop}
\ee
where $k$ is the momentum of the photon in the loop and $\sigma(k,p)$
is the appropriate, IV self-energy integrand, a number of which are
defined below.  The individual FV and IV terms in \eq{eq:DSig1loop}
are generally UV and possibly IR divergent. Thus, individually they
should be regularized, e.g.\ with dimensional regularization. However,
on shell the IV integral is IR finite and in finite
volume, the sums are IR finite because the FV formulations of QED that
we consider are regulated by the space or spacetime volume. Moreover,
for large $k^2$, the functions $k^3\sigma(k,p)$ that arise in
\eq{eq:DSig1loop} are strictly monotonic. Therefore, the proof of
Cauchy's integral criterion \cite{wiki:cauchytest} guarantees that the
difference of the FV sums and IV integrals is also UV finite. Thus, to
compute the FV corrections at one loop, no explicit regularization is
required.

In \eq{eq:DSig1loop}, the information about the topology of the finite
volume and the specific formulation of QED is contained in the
definition of $\sumint_k^\prime$. In addition to the case of $\T^4$
already discussed, we will also consider the four-cylinder
$\R\times\T^3$. This is a useful intermediate step computationally,
because it allows single particle propagators to develop a pole at the
particle's energy. 

For the cases of interest here, we have the following definitions for
$\sumint_k^\prime$:
\begin{itemize}

\item $\qedL$ on $\R\times\T^3$:
\be
\sumint_k^\prime \equiv \int_{-\infty}^{+\infty}\frac{dk_0}{2\pi}\frac1{L^3}
\sum_{\vec{k}\in\BZL^{3*}}
\ ,
\labell{eq:QEDLonRT3}
\ee
with $\BZL^{3*}\equiv\TBZL^{3*}$ and the star, as usual, indicates the
removal of the zero element;

\item $\qedL$ on $\T^4$:
\be
\sumint_k^\prime \equiv \frac1{TL^3}
\sum_{k_0\in\BZT}\sum_{\vec{k}\in\BZL^{3*}}
\ ,
\labell{eq:QEDLonT4}
\ee
with $\BZT\equiv\TBZT$;

\item $\qedTL$ on $\T^4$:
\be
\sumint_k^\prime \equiv \frac1{TL^3}
\sum_{k_\mu\in\BZTL^{4*}}
\ ,
\labell{eq:QEDTLonT4}
\ee
where $\BZTL^{4*}\equiv[\TBZT\times\TBZL^3]^{*}$.

\end{itemize}

The last ingredient of a general nature, needed to study the FV
corrections in the three cases of interest, is the integrand of the
self-energy $\sigma_J(k,p)$ for fermions ($J=1/2$) and bosons
($J=0$). These are obtained from the usual one-loop spinor and scalar
self-energy Feynman diagrams, yielding the following expressions:
\bea
        \sigma_{\frac{1}{2}}(k,p) &=& (2i\slashed{p}+4m)\sigma_{S_1}(k,p)+
        2i\slashed{\sigma}_{S_2}(k,p)
\labell{eq:sig12}
\\
        \sigma_0(k,p) &=& 4\sigma_T(k)-
\sigma_{S_0}(k,p)-4p^2\sigma_{S_1}(k,p)-4p_{\mu}
        \sigma_{S_2,\mu}(k,p)
\ ,
\labell{eq:sig0}
\eea
with,
\be
\begin{array}{cc}
\sigma_T(k) = \frac{1}{k^2}\ , & \sigma_{S_0}(k,p) = \frac{1}{[(p+k)^2+m^2]}\ ,   \\  
\sigma_{S_1}(k,p) = \frac{1}{k^2[(p+k)^2+m^2]}\ , & \sigma_{S_2,\mu}(k,p) =\frac{k_{\mu}}{k^2[(p+k)^2+m^2]}
\ .
\end{array}
\labell{eq:sigTSi}
\ee

In the following subsection we provide a brief description of the
methods used to obtain the FV corrections in inverse powers of the
volume for the three formulations of FV QED described above. We summarize
the results and discuss their consequences in \sec{sec:qedana_res}.

\subsection{Computation of finite-volume corrections in various QED formulations}

\subsubsection{Finite-volume corrections in $\qedL$ on 
$\R\times\T^3$}

$\qedL$ on $\R\times\T^3$ is not a setup
that can be considered directly in lattice simulations, as it
describes a spacetime with an infinite time direction. However, it is
a useful first step for computing FV corrections to masses
analytically. These corrections are obtained from
\eqs{eq:Dmn}{eq:DSig1loop} with the FV self-energy sum defined through
\eq{eq:QEDLonRT3}. To evaluate the resulting expressions and obtain
an asymptotic expansion in powers of $1/L$, we apply the Poisson
summation formula to the sum over the three-momentum $\vec{k}$,
subtracting appropriately the $\vec{k}=0$ modes. Then, using
techniques from \cite{Hasenfratz:1989pk,Hayakawa:2008an} and carrying
out the asymptotic expansions to the end, up to exponentially small
corrections in $mL$, we obtain:
\bea
\label{eq:qedana}
\Deltaop{\qed\to\qedL}{\R^4\to\R\times\T^3}
\begin{Bmatrix}
\Sigma_T(L)\\
\Sigma_{S_0}(im,L)\\
\Sigma_{S_1}(im,L)\\
\Sigma_{S_2,\mu}(im,L)\\
\end{Bmatrix}
&=&
\left[\sum_{\vec{x}\in L\Z^{3*}}-\frac1{L^3}\int d^3x\right]
\int\frac{d^4k}{(2\pi)^4}
\begin{Bmatrix}
\frac{1}{k^2} \\
\frac{1}{2imk_0+k^2} \\
\frac{1}{k^2[2imk_0+k^2]} \\
\frac{k_{\mu}}{k^2[2imk_0+k^2]}\\
\end{Bmatrix}
e^{i\vec{k}\cdot\vec{x}}\nonumber\labell{eq:QEDLonRT3poisson}\\
&\underset{L\to+\infty}{\sim}&
\begin{Bmatrix}
-\frac{\kappa}{4\pi L^2}\\
-\frac{1}{2 m L^3}\\
-\frac{\kappa}{16\pi mL}+\frac{1}{8m^3L^3}\\
im\delta_{\mu 0}\left( \frac{\kappa}{8\pi m^2L^2}-\frac{1}{4m^3L^3} \right)
\end{Bmatrix},
\labell{eq:fvintegrals}
\eea
where
\bea
\kappa &\equiv& \int_0^\infty \frac{d\lambda}{\lambda^{3/2}}\left\{\lambda^{3/2}+1-
\left[\mathcal{\theta}_3(0,e^{-\frac\pi\lambda})\right]^3\right\}
\labell{eq:kappa}
\nn\\
&=&
2.837297(1)\ ,
\eea
and where $\mathcal{\theta}_3(u,q)=\sum_{n\in\mathbb{Z}}
q^{n^2}e^{i2nu}$ is a Jacobi theta function.  In \eq{eq:fvintegrals},
$\Sigma_i(im,L)$ is the on-shell self energy corresponding to $\sigma_i(k,im)$,
$i=T,\cdots,S_{2,\mu}$, and the notation
$\Deltaop{\qed\to\qedL}{\R^4\to\R\times\T^3}$ indicates that these
corrections must be added to the relevant quantity determined in
$\qed$ on $\R^4$ (i.e.\ standard IV QED) to obtain the quantity
appropriate for $\qedL$ on $\R\times\T^3$. A similar notation is used
below for other corrections, with a meaning which is a straightforward
generalization of the one described here.

\subsubsection{Finite-volume corrections in $\qedL$ on 
$\T^4$}

The FV corrections to the
mass of a point particle are obtained from \eqs{eq:Dmn}{eq:DSig1loop}
with the FV self-energy sum defined through
\eq{eq:QEDLonT4}. Instead of performing an asymptotic expansion for
$T,L\to\infty$ directly on this expression, it is easier to compute
the corrections to the results obtained for $\qedL$ on
$\R\times\T^3$ that result from compactifying the time direction to
a circle of circumference $T$. In that case, instead of the expressions in
\eq{eq:DSig1loop}, we must compute:
\bea
\Deltaop{\qedL\to\qedL}{\R\times\T^3\to\T^4}\Sigma(im,T,L) 
& = & \left[\frac{1}{T}\sum_{k_0\in\BZT}-\int\frac{dk_0}{2\pi}\right]
\sum_{\vec{k}\in\BZL^{3*}}\sigma(k,im)
\labell{eq:DmnRT3toT4}\\
&=&\sum_{x_0\in T\Z^*}\sum_{\vec{k}\in\BZL^{3*}}\int\frac{dk_0}{2\pi}
\sigma(k,im)e^{ik_0x_0}
\ ,
\labell{eq:DmnRT3toT4Poisson}
\eea
where, again, we have used Poisson's summation formula. Inspection
of \eq{eq:sigTSi} indicates that the functions
$\sigma(k,im)$ have no poles on the real
$k_0$-axis, are infinitely differentiable and all of their derivatives
are integrable. Therefore, their Fourier transform in
\eq{eq:DmnRT3toT4Poisson} vanishes faster than any power of $1/T$ as
$T\to\infty$. This means that the FV corrections to the on-shell
self-energy in $\qedL$ on the four-torus of dimensions $T\times
L^3$, are the same as those on the four-cylinder $\R\times\T^3$, up to
corrections that vanish faster than any inverse power of $T$ , i.e.:
\be
\Deltaop{\qed\to\qedL}{\R^4\to\T^4}\Sigma(im,T,L) \underset{T,L\to+\infty}{\sim} \Deltaop{\qed\to\qedL}{\R^4\to\R\times\T^3}\Sigma(im,L)
\ .
\labell{eq:DRT3toT4res}\ee

\subsubsection{Finite-volume corrections in $\qedTL$ on 
$\T^4$}

In this setup, the FV corrections to the self-energy of a point
particle are obtained from \eq{eq:DSig1loop} with the FV sum defined
through
\eq{eq:QEDTLonT4}. As in the previous section, instead of performing
a $T,L\to\infty$ asymptotic expansion directly on this expression, we
compute the corrections to the results obtained for $\qedL$ on
$\T^4$ which we obtain from reinstatement, as dynamical variables, the
photon field modes $\tilde A_\mu(k_0,\vec0)$ with $k_0\ne 0$. These
corrections require computing:
\be
\Deltaop{\qedL\to\qedTL}{\T^4\to\T^4} \Sigma(p,T,L) 
= \frac{1}{TL^3}\sum_{k_0\in\BZT^*,\,\vec{k}=\vec0}
\sigma(k,p)
\ ,
\labell{eq:DmnLtoTL}
\ee
for the various self-energy integrands $\sigma(k,p)$ of
\eqsmany{eq:sig12}{eq:sigTSi}.

The functions which appear in $\sigma(k,p)$, for $\vec{k}=\vec0$, are
rational functions of $k_0$. There are known systematic methods to sum
series of such functions. These involve performing partial fraction
decompositions and then exploiting the properties
of the polygamma functions to sum the individual terms in these
decompositions. Using these methods, we obtain:
\be
\Deltaop{\qedL\to\qedTL}{\T^4\to\T^4}
\begin{Bmatrix}
\Sigma_T(T,L)\\
\Sigma_{S_0}(im,T,L)\\
\Sigma_{S_1}(im,T,L)\\
\Sigma_{S_2,\mu}(im,T,L)\\
\end{Bmatrix}
\underset{T,L\to+\infty}{\sim}
\begin{Bmatrix}
\frac{T}{12L^3}\\
\frac{\coth\left({mT}\right)}{4mL^3}-\frac{1}{4m^2TL^3}\\
\frac{T}{48m^2L^3}-\frac{\coth(mT)}{16m^3L^3}+\frac{1}{16m^4TL^3}\\
-2im\delta_{\mu 0}\,\Sigma_{S_1}(im,T,L)
\end{Bmatrix}
\labell{eq:DLtoTLres}
\ee
which are the corrections that must be added to the self-energy
contributions in $\qedL$ on $\T^4$ to obtain the FV contributions in
$\qedTL$ on $\T^4$.

\subsection{Results for finite-volume corrections to the pole mass and consequences for the various QED formulations}
\labell{sec:qedana_res}

In this subsection we combine the results of the previous subsection
to obtain the FV corrections to the pole masses of point spinor and
scalar particles for two versions of FV QED of interest for lattice
calculations:
\begin{itemize}
\item $\qedL$ on $\T^4$, which is the formulation used in the present study;
\item $\qedTL$ on $\T^4$, which is the formulation used in previous lattice studies of isospin breaking effects.
\end{itemize}
While some of the details of the results obtained in this section are
specific to point particles, the general conclusions also carry over
to the case of composite particles, which is discussed
in \sec{se:fvol}.

\subsubsection{Finite-volume corrections in $\qedL$ on 
$\T^4$}

$\qedL$ on $\T^4$ is the formulation used in the present study of
isospin breaking effects. Combining the results
of \eqs{eq:fvintegrals}{eq:DRT3toT4res} and putting everything
together, we find that the mass of a point-like
fermion of spin $1/2$, of charge $q$ in units of $e$, on the
four-torus $\T^4$ of dimensions $T\times L^3$ in $\qedL$, is at
one-loop, in terms of its infinite-volume mass $m$:
\be
m_{\frac12}(T,L) \underset{T,L\to+\infty}{\sim} 
m\left\{1-q^2\alpha\left[\frac{\kappa}{2mL}\left(1+\frac{2}{mL}\right)
-\frac{3\pi}{(mL)^3}\right]\right\}
\ , 
\labell{eq:Dm12QEDLonT4}
\ee
up to terms which are exponentially suppressed in $mL$ and terms which
fall faster than any power in $1/(mT)$, with $\kappa$ given
in \eq{eq:kappa}. Similarly, the FV corrections to the mass of a
point-like boson of spin $0$ are, in terms of its infinite-volume mass
$m$:
\be
m_0^2(T,L) \underset{T,L\to+\infty}{\sim} m^2\left\{1-q^2\alpha
\left[\frac{\kappa}{mL}\left(1+\frac{2}{mL}\right)\right]\right\}
\ .
\labell{eq:Dm0QEDLonT4}
\ee

Four important comments are in order. The first is that the finite-volume
pole masses in both cases have a well defined $T,L\to\infty$ limit and
converge onto their infinite-volume counterparts. The second is that
the coefficient of the leading $1/L$ and $1/L^2$ corrections to the
mass $m$ of a particle of charge $qe$ is the same for spin-$1/2$
fermions and spin-$0$ bosons at $O(\alpha)$. In \sec{se:fvol} we show
that these coefficients are always the same, independent of the spin
and point-like nature of the particle: they are fixed by QED
Ward-Takahashi identities. Moreover, as suggested
in \cite{Portelli:2010yn} and worked out explicitly in \cite{Davoudi:2014qua},
the leading $1/L$ term is the FV correction to the classical,
electrostatic potential of a point charge on $\T^3$, with the spatial
zero-modes removed from Gauss' law. The third comment is that the dimensionless,
relative FV corrections must be functions of the only dimensionless
parameter, $mL$, in the two theories considered. This will no
longer be the case when we consider physical mesons and baryons, as these
particles are not point-like and therefore have relevant scales other
than their mass. The final remark is that we find a coefficient for
the $1/(mL)^3$ term in \eq{eq:Dm12QEDLonT4} which is twice the one
found in \cite{Davoudi:2014qua}, when the result for composite
fermions in that paper is reduced to the point-like case. As shown
in \sec{se:qednum}, this factor of $2$ is confirmed by direct
simulation of $\qedL$ on $\T^4$.

\subsubsection{Finite-volume corrections in $\qedTL$ on 
$\T^4$}

The setup considered in this section, $\qedTL$ on the four-torus
$\T^4$ of dimensions $L^3\times T$, is the one used in all previous
studies which include QED corrections to hadronic observables in
lattice
QCD\cite{Duncan:1996xy,Duncan:1996be,Blum:2007cy,Portelli:2010yn,Blum:2010ym,Ishikawa:2012ix,Aoki:2012st,deDivitiis:2013xla,Borsanyi:2013lga}. As
discussed in \sec{se:phact}, it violates reflection positivity. Here
we show that it has another problem: it does not have a well defined
$T\to\infty$ limit for fixed $L$. It is these reasons which have led
us to choose to simulate $\qedL$ instead of $\qedTL$ for the precision
computation presented in the present paper.

The finite-volume corrections to the masses of point particles in
$\qedTL$ on $\T^4$ are obtained by adding, to those in $\qedL$ on
$\T^4$ (\eqs{eq:Dm12QEDLonT4}{eq:Dm0QEDLonT4}), the corrections on the
self-energy components determined in \eq{eq:DLtoTLres}. This yields
the following result for the mass of a spin $J=1/2$ point-particle,
of charge $q$ in units of $e$, in $\qedTL$ on the
four-torus of dimensions $T\times L^3$, in terms of its
infinite-volume counterpart, $m$:
\bea
m_{\frac12}(T,L) &\underset{T,L\to+\infty}{\sim}& 
m\left\{1-q^2\alpha\left[\frac{\kappa}{2mL}
\left(1+\frac{2}{mL}\left[1-\frac{\pi}{2\kappa}\frac{T}{L}\right]\right)
\right.\right.\nn\\
&&\left.\left.-\frac{3\pi}{(mL)^3}\left[1-\frac{\coth(mT)}{2}\right]-
\frac{3\pi}{2(mL)^4}\frac{L}{T}\right]\right\}
\ , 
\labell{eq:Dm12QEDTLonT4}
\eea
up to terms which are exponentially suppressed in $mL$ and terms which
fall faster than any power in $1/(mT)$, with $\kappa$ given
in \eq{eq:kappa}. Similarly, the FV corrections to a point-like boson
of spin $0$ are, in terms of the infinite-volume mass $m$:
\bea
m_0^2(T,L) &\underset{T,L\to+\infty}{\sim}& m^2\left\{1-q^2\alpha
\left[\frac{\kappa}{mL}\left(1+\frac{2}{mL}\left[1-\frac{\pi}{2\kappa}\frac{T}{L}\right]\right)\right]\right\}
\ .
\labell{eq:Dm0QEDTLonT4}
\eea

A number of important remarks about these results deserve to be
made. We begin by considering $T/L$ as being $O(1)$, as it is usually
in lattice simulations.  The first remark is that the leading $1/L$
and $1/L^2$ contribution are identical for both spins, as was the case
in $\qedL$. However, only the
$1/L$ terms here are equal with those found in $\qedL$. The
reinstatement of the spatially-uniform photon modes for $k_0\ne 0$
reduces the coefficient of the subleading $1/L^2$ contributions for
$T\sim L$, compared to what it is in $\qedL$.

It should also be noted that both masses acquire new $T$-dependent,
$1/L^3$ and $1/L^4$ contributions, which remain
under control even when $T\gg L$. And, as in the case of $\qedL$, the
dimensionless, relative, FV mass corrections
of \eqs{eq:Dm12QEDTLonT4}{eq:Dm0QEDTLonT4} can be written in terms
only of dimensionless quantities. However, while in $\qedL$ there is
only one dimensionless variable, $mL$, here there is another, the
aspect ratio of $\T^4$, $\xi=T/L$. This will no longer be the case
when we consider physical mesons and baryons, as these particles are not
point-like and therefore have relevant scales other than their mass.

The problem in $\qedTL$ on the four-torus arises if one considers the
limit $T\to\infty$ for fixed $L$. In that case the $1/L^2$ corrections
blow up linearly in $T/L$. In particular, this means that the
conventional mass extraction procedure, which relies on examining the
asymptotic time behavior of Euclidean propagators, is not
well-defined at finite $L$. This justifies, for the precision study presented in
this paper, our choice of working with $\qedL$ instead of $\qedTL$
that was used in previous studies. It should be noted, however, that
this problem only arises at $O(1/L^2)$ and remains mild for typical
$T/L$ used in lattice computations. Thus, for lower precision studies
with $T/L\sim 1$ such as in \cite{Borsanyi:2013lga}, where no
sensitivity to terms of higher order than a fitted $1/L$ was found, the effect
of using $\qedTL$ will not significantly distort results.

\section{Finite-volume and lattice-spacing corrections in QED: numerical investigations}
\label{se:qednum}

The results of this section serve to verify the implementation of the QED part
of our simulation, which is one of the novel features of this work. The infinite volume
and continuum extrapolations will be shown to be under control. Both
extrapolations from the numerical data yield results that are consistent with
analytical calculations.

Our QED setup is obtained from the full case by setting the gluon fields to
$U=1$. Since we focus on the order $e^2$ behavior, we will use the quenched
QED approximation in this illustrative section, which is correct up to $e^4$
effects. We take $e=\sqrt{4\pi/137}$ and a fermion charge of $q=1$ unless
stated otherwise. In some cases we switch off the clover improvement and/or the
photon field smearing.  

The field configuration generation in the absence of fermion loops can be done by
a heatbath algorithm.  Using a normal Gaussian distributed random vector
$\eta_\mu$, a Feyman-gauge field configuration with the right distribution is
given by $A_{\mu,k}=|\hat{k}|^{-1}\eta_{\mu,k}$. This is
then fixed to Coulomb-gauge by the operator in \eq{eq:pc}. 

\begin{figure}[H]
\centering
\includegraphics{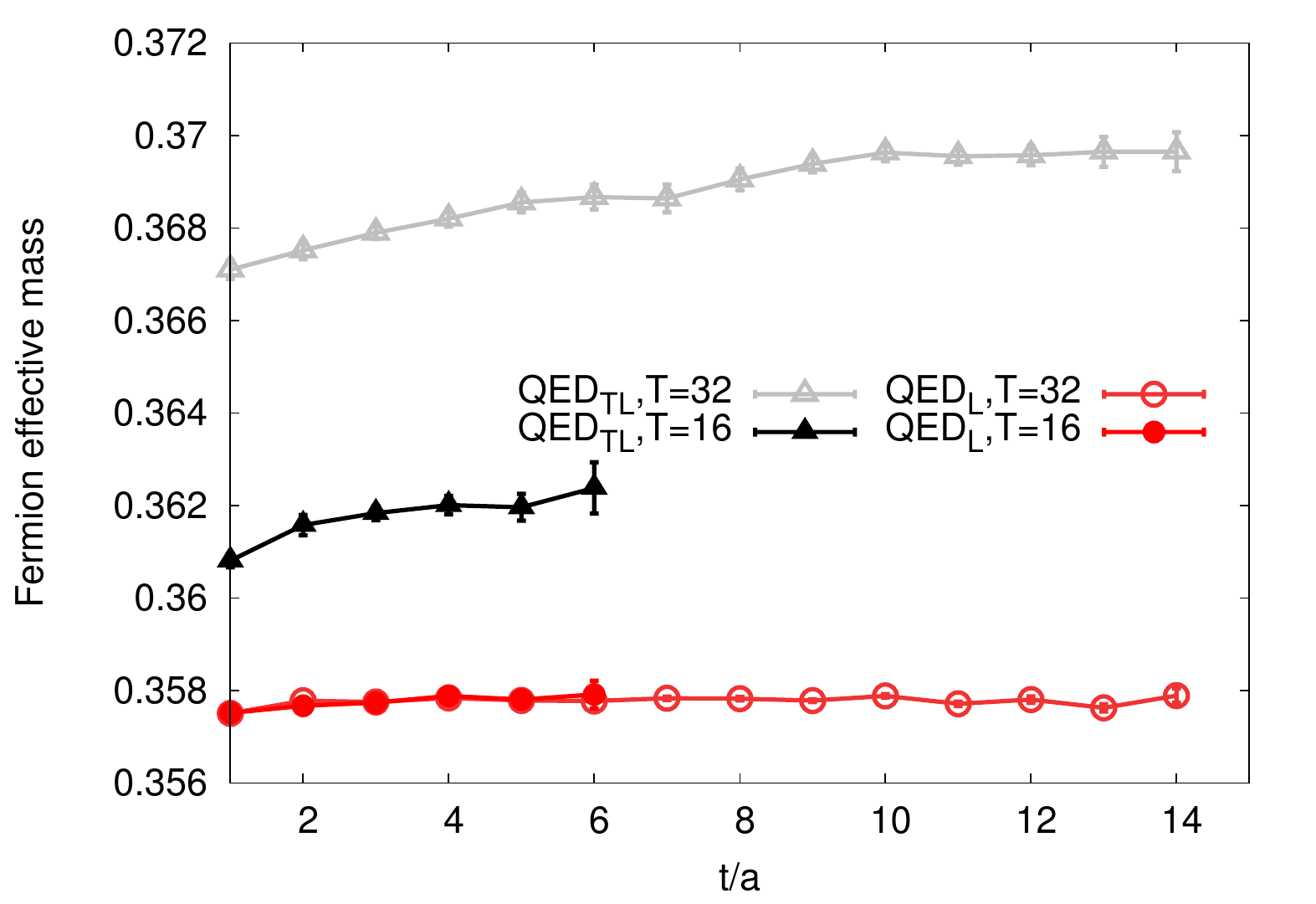}
\caption{\label{fi:qednum0}
Quark effective masses in QED using different zero-mode subtractions. The four-dimensional
zero-mode subtraction ($\qedTL$) produces masses that depend on $T$. The time-slice by time-slice removal ($\qedL$) makes
effective masses well behaved.
}
\end{figure}

Let us first take a look at the effective mass of a fermion in the two different
realizations of finite volume QED: $\qedTL$ and $\qedL$ (see 
\fig{fi:qednum0}). The difference is in the zero-mode removal prescription.  The
following parameters were used: spatial box size $L=4$, bare Wilson mass
$m=0.4$, no clover improvement, no smearing.  In $\qedTL$ no clear plateaus can
be seen, moreover increasing the time extent also increases the mass values.
This is in complete agreement with the analytical finding of 
\sec{se:qedana}: the mass diverges in the infinite time extent limit, if the spatial
box size is kept fixed. This of course makes the conventional mass extraction
procedure, which examines the asymptotic behavior of the propagator,
ambiguous in this scheme. If instead we work with the zero-mode prescription
of $\qedL$, we get masses that are insensitive to the time extent of the
lattice, as predicted.

\begin{figure}[H]
\centering
\includegraphics{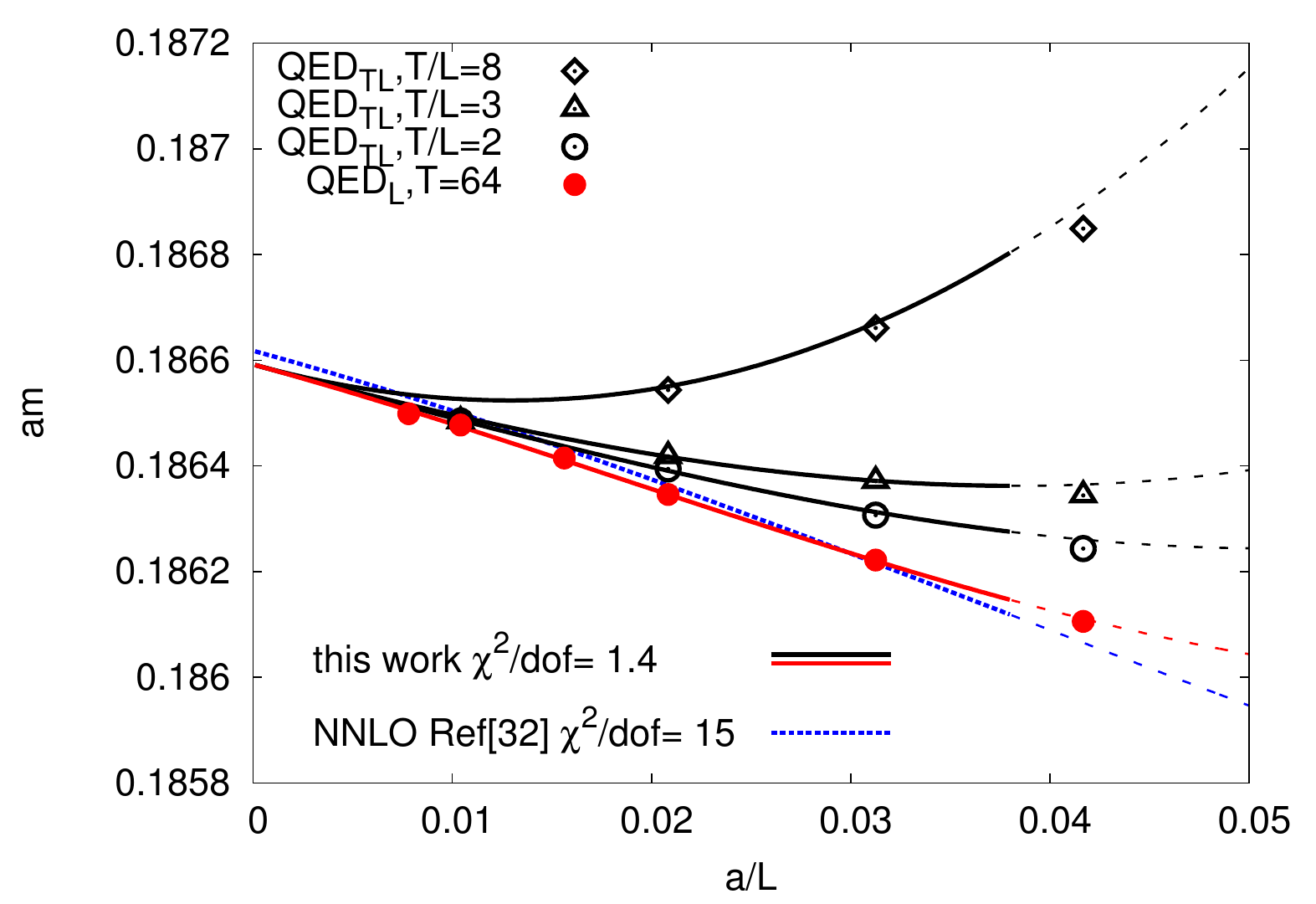}
\caption{\label{fi:qednum1}
Finite-size dependence of the fermion mass in two different realizations of finite-volume QED,
that differ in the zero-mode subtraction.
The open/filled symbols are obtained using $\qedTL$/$\qedL$ prescription.
The curves
are the analytical formulae from \eqs{eq:Dm12QEDLonT4}{eq:Dm12QEDTLonT4}. The dotted
curve is the NNLO formula from Ref.\ \cite{Davoudi:2014qua}.
}
\end{figure}

Now let us consider the finite size dependence of the fermion mass (see
\fig{fi:qednum1}). For the plot we used a bare Wilson mass of $m=0.2$, clover
improvement and smearing. The points with filled symbols are obtained in the
$\qedL$ formulation.  We also extract masses using data in the $\qedTL$ setup,
though identification of the plateaus
is sometimes unclear, as discussed above.  These are the open symbols on the plot and correspond to three
different aspect ratios.  The masses change with $T$ at a fixed spatial volume.
We fit the results of $\qedL$ and $\qedTL$ together to the analytical formulae of
\eqs{eq:Dm12QEDLonT4}{eq:Dm12QEDTLonT4} with a single fit parameter, which
is the infinite volume fermion mass. A good quality fit can be achieved, if we
restrict the boxes to $a/L<0.04$. The results are the solid lines on the plot.
The small disagreements for smaller boxes can probably be explained by
exponentially small finite size corrections, which are neglected in the
formulae. 
The data has sufficient precision to determine the coefficient of the
$1/L^3$ term. It strongly favors the one that we find by analytic
calculation in \eq{eq:Dm12QEDTLonT4} over the one obtained by reducing the
composite-particle result of \cite{Davoudi:2014qua} to the point-particle
case.~\footnote{We perform this reduction by setting the charge radius
      and all other coefficients which reflect the composite nature of
  the particle to zero.} Fitting the numerical results to the formula
of \cite{Davoudi:2014qua} leads to a $\chi^2/\mathrm{dof}=1.4$. A fit of the
reduced, NNLO expression of \cite{Davoudi:2014qua} to the $\qedL$ points gives a $\chi^2/$dof$= 15$.
These fits are shown in \fig{fi:qednum1} with solid/dotted lines.
Including also the small
volumes in the analysis (dashed lines) the $\chi^2/$dof is increased by a factor of $2.5$ for \eq{eq:Dm12QEDTLonT4},
whereas the formula of \cite{Davoudi:2014qua} provides a $\chi^2/$dof well over 100.

\begin{figure}[H]
\centering
\includegraphics{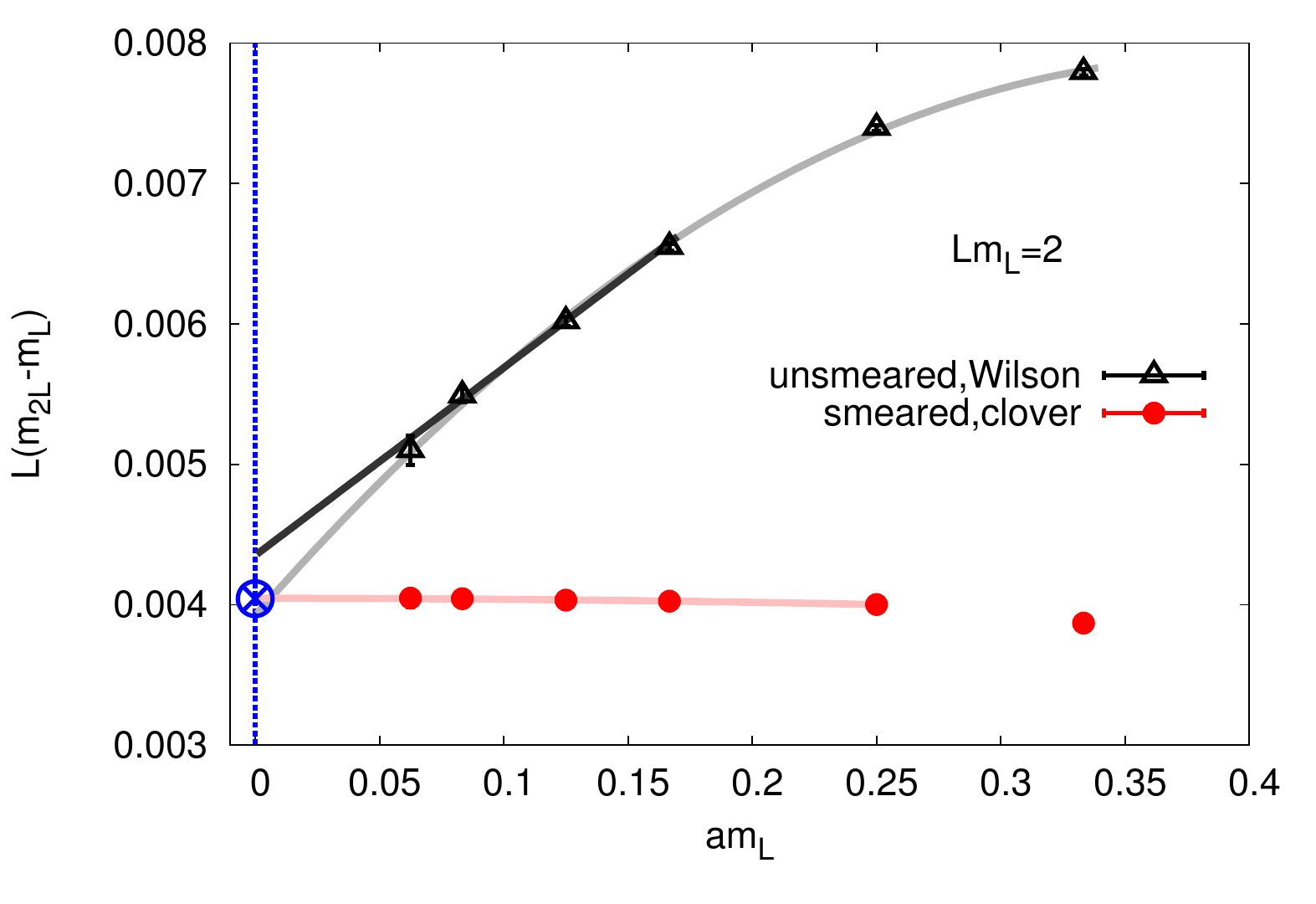}
\caption{\label{fi:qednum2}
Continuum scaling of a mass difference in $\qedL$ using two different
lattice actions. The crossed circle in the continuum limit is the analytical result from the expressions of \sec{se:qedana} (see text).
The fitted curves are described in the text.
}
\end{figure}

Now let us turn our attention to the lattice artefacts of our QED lattice
action.
In order to carry out a lattice spacing scaling study, we need
two observables: one which sets the lattice spacing and another which has a
non-trivial continuum limit\footnote{
Since we work
in the quenched approximation in this section, the triviality issue is absent. This is because the renormalized coupling 
equals the bare coupling and it can be kept fixed for arbitrary small lattice spacings.}.
For the first the obvious choice is the fermion
mass, in a given finite box. We denote it $m_L$.  For the second we take the
finite volume correction: the difference of $m_L$ to the mass measured on a
lattice with twice the spatial extent, denoted by $m_{2L}$. We set the box size
as $Lm_L=2$ and work in the $\qedL$ formulation.  The result is shown in
\fig{fi:qednum2}. The filled/open points are obtained with/without clover
improvement and smearing. In the unimproved case we use two functions to
extrapolate into the continuum: one is linear and the other is
linear-plus-quadratic in the lattice spacing.  For the improved case a
quadratic function is plotted, but the data is even consistent with a constant
after ignoring the result from the coarsest lattice.  A continuum value can
also be obtained analytically using the results of the previous section.
At a mass of $Lm_L=2$ the exponential corrections cannot be neglected. Thus
we compute the $L(m_{2L}-m_L)$ using \eq{eq:Dmn} and evaluating 
the relevant integrals and sums in \eq{eq:qedana} numerically instead
of using the asymptotic formulae.
This is the crossed circle on the plot.
Both the unimproved and the
improved results agree with the analytical formula, though the unimproved has a
much larger systematic error coming from the continuum extrapolation.

Finally in order to further support our QED setup, we compare the infinite volume extrapolated fermion masses at finite
lattice spacing to analytical results from the literature. In 
\cite{Mertens:1997wx} Mertens, Kronfeld and El-Khadra calculated the pole
position of the fermion propagator for arbitrary Wilson mass, clover coefficient
and to leading order in $e^2$. For the comparison we used the $\qedL$ formulation,
clover improvement and unsmeared photons. The bare Wilson mass was set to
$am=0.2$. We carried out the infinite volume extrapolation using the analytical
finite volume formula, \eq{eq:Dm12QEDLonT4}. There is a small discrepancy
at the standard coupling value, $e=\sqrt{4\pi/137}$. To demonstrate, that it is an $O(e^4)$ effect,
we carried out simulations at two other coupling values. The infinite volume
extrapolated results are recorded in \tab{ta:qednum}. The difference to
the analytical result is consistent with being an $e^4$ correction.

\section{Finite-volume effects for composite particles}
\label{se:fvol}

As discussed in \sec{se:qedana}, finite-volume (FV) effects are
particularly important, because of the presence of the massless
photon. While in that section we computed these effects for point-like
scalar and spinor particles, here we generalize the discussion to
composite particles, which can be hadrons, nuclei, etc. We
consider only the QED formulation that was actually used in our
numerical work, namely $\qedL$ on the four-torus $\T^4$ of dimensions
$T\times L^3$ and use the fact that it differs from $\qedL$ on
$\R\times \T^3$ by corrections in $T$ which fall of faster than any
power in $1/T$ (see \sec{se:qedana}).

As shown in \sec{se:qedana}
(\eqs{eq:Dm12QEDLonT4}{eq:Dm0QEDLonT4}), the leading and
next-to-leading FV corrections to the mass of a point particle of
charge $qe$ in scalar and fermion QED are identical at
$\mathcal{O}(\alpha)$:
\be
m(T,L) \underset{T,L\to+\infty}{\sim} 
m\left\{1-q^2\alpha\frac{\kappa}{2mL}\left[1+
\frac{2}{mL}\right]+\ord{\frac{\alpha}{L^3}}\right\}
\labell{eq:Dmuniversal}
\ ,\ee
with $\kappa$ given in \eq{eq:kappa}. Here we show that both
corrections remain unchanged for composite particles. This result
follows from the QED Ward-Takahashi identities (WTIs) and from the
assumptions that the photon is the only massless asymptotic state, and
that the charged particle considered is stable and non-degenerate in
mass with any other particle which shares its quantum numbers.

While the project described in the present paper was being finalized,
a determination of FV corrections to the masses of composite particles
was presented in \cite{Davoudi:2014qua}, up to $\ord{1/L^4}$. In that
paper, the authors use an elegant approach based on heavy-hadron
effective field theories. Nevertheless, because of the importance of
the infinite-volume (IV) extrapolation in studies involving QED and because
the precision of our results depends critically on knowing the value
of the coefficients of the $1/L$ and $1/L^2$ corrections, we have
chosen to outline our independent derivation of \eq{eq:Dmuniversal} in the
present section. 
Although we work out the corrections only up to $\ord{1/L^2}$, our approach can be
considered more general, as it is based on general field theoretical
considerations and not on a specific effective theory.

Note that \eq{eq:Dmuniversal} implies that, to $\ord{1/L^2}$ at least,
composite, neutral particles receive no FV
corrections. According to \cite{Davoudi:2014qua}, these only appear at
$\ord{1/L^4}$ for such particles.

\begin{figure}[H]
\includegraphics*[bb=45 652 562 738]{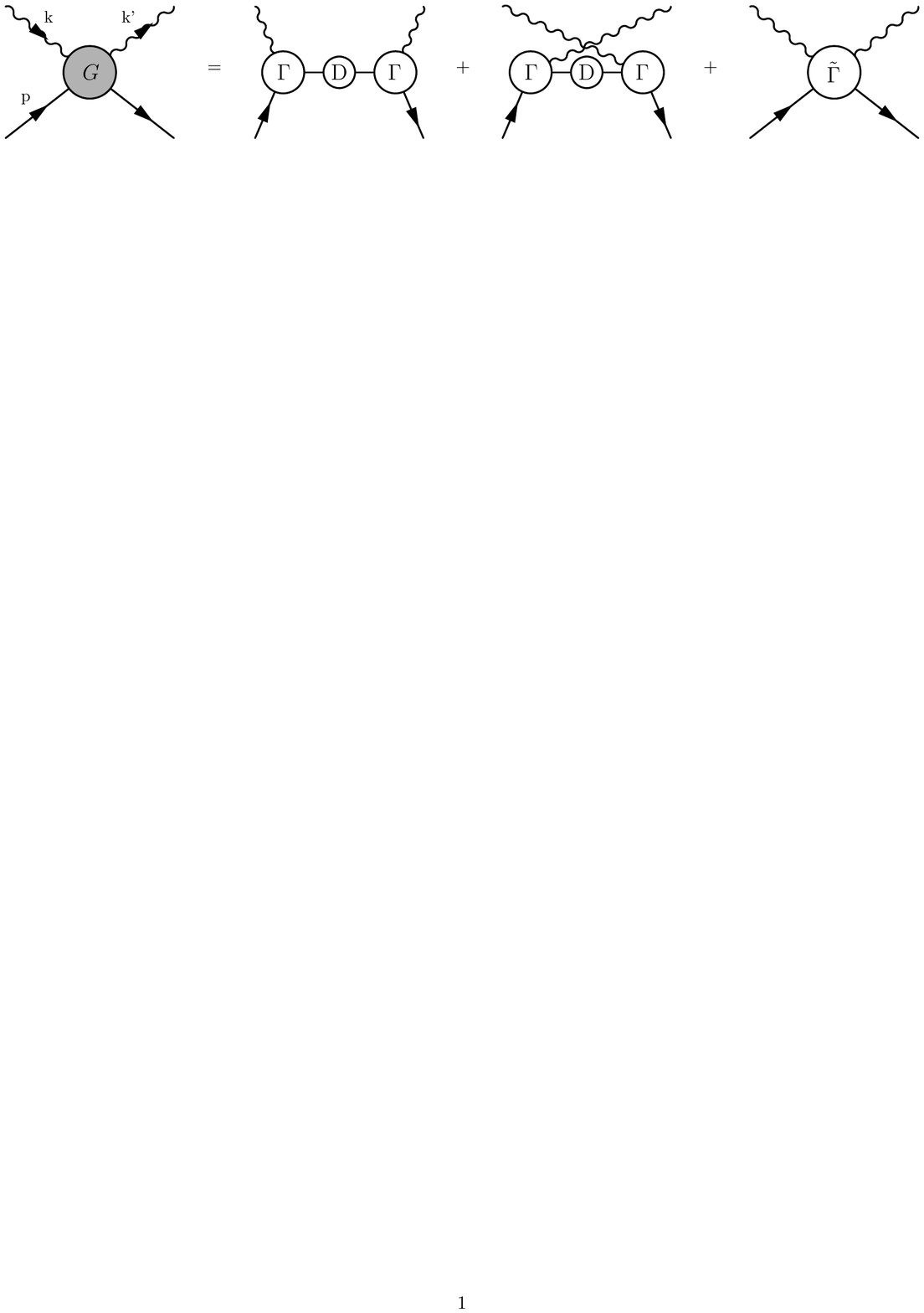}
\caption{\label{fi:feynman}Decomposition of the four-point vertex into 1PI parts.}
\end{figure}

Here we consider a scalar particle of mass $m$ and charge $qe$. The fermionic
case is entirely analogous, but is technically more involved because of spin.
As in the point-particle case, power-like FV corrections to the mass of
composite particles arise from the exchange of a massless photon around the
spacetime torus.  The modification in the case of composite particles comes
from the fact that the photons couple to a state which has internal structure.
The coupling, that enters in the self-energy calculation, is described by the
connected scalar-scalar-photon-photon vertex $G(k';p,k)$, with the momentum
of the outgoing scalar equal to $p+k-k'$. At $\ord{\alpha}$ it can be decomposed
as shown in \fig{fi:feynman}.  $D(p)$ is the full, re-summed scalar
propagator. $\Gamma_\mu(p,k)$ is the one-particle irreducible (1PI)
scalar-scalar-photon, three-point vertex function, with the momentum of the
outgoing scalar equal to $p+k$.  $\tilde{\Gamma}_{\mu\nu}(k';p,k)$ is the 1PI
scalar-scalar-photon-photon, four-point vertex function, with the momentum of
the outgoing scalar equal to $p+k-k'$. These functions contain no photon lines and
they are $\ord{\alpha^0}$. They are all infinite-volume,
renormalized quantities.
They satisfy the following WTIs:
\bea
k_\mu\Gamma_\mu(p,k)/q&=& D^{-1}(p+k)- D^{-1}(p)
\labell{eq:3ptWTI}\\
-k_\mu k_\nu\tilde{\Gamma}_{\nu\mu}(k;p,k)/q^2&=& D^{-1}(p+k)+ D^{-1}(p-k)
\ ,\labell{eq:4ptWTI}
\eea
where again $q$ is the charge of the particle in units of $e$.

The FV corrections to the mass are obtained by taking the difference of the
particle's on-shell self-energy in finite and infinite volumes. 
For a scalar particle at $\ord{\alpha}$, this yields:
\begin{gather}
\labell{eq:dm2fvmaster}
\Delta m^2(T,L) \equiv m^2(T,L)-m^2= -\frac{e^2}{2} \left[\sumint_k^\prime-\int\frac{d^4k}{(2\pi)^4}\right] \frac{1}{k^2}G(k;p=im,k)=\\
 = -\frac{e^2}2\left[\sumint_k^\prime-\int\frac{d^4k}{(2\pi)^4}\right]\frac{1}{k^2}\{2\Gamma_\mu(p+k,-k) D(p+k)\Gamma_\mu(p,k)\nn
 +\tilde{\Gamma}_{\mu\mu}(k;p,k)\}_{p=im},
\end{gather}
where the sum is the one defined by \eq{eq:QEDLonRT3}.  Since the
contributions to the pole mass are gauge independent, we work in the simpler
Feynman gauge.  

In \eq{eq:dm2fvmaster} an important step has already been
taken: the propagator $D$ and the vertices $\Gamma_\mu$ and
$\tilde{\Gamma}_{\mu\nu}$ have been replaced by their IV counterparts. This is
because, for the kinetics of the self-energy diagrams these
functions differ only by terms that fall off exponentially with the size of the
box.  For the propagator of a stable particle with an inflowing energy smaller
than that of any possible on-shell, multi-particle state, this was shown by
L\"uscher \cite{Luscher:1985dn}. In the same paper the analiticity of the
three-point and four-point 1PI vertex functions and the associated
Feynman-integrands is proven (Theorem 2.3), assuming again, that the inflowing
energy is smaller than that of any possible multi-particle states. The FV
vertices are obtained by replacing the momentum integrals, which
appear in the loop expansions of their IV counterparts, by momentum
sums. The analyticity proof applies also in FV.
Combined with the well known properties of Fourier transforms, this
guarantees that the FV corrections to the vertex functions,
computed using Poisson's formula, fall off faster than any inverse
power of size of the box.
The same arguments apply to the 1PI
vertices $\Gamma_\mu$ and $\tilde{\Gamma}_{\mu\nu}$ considered here.

No regularization is required in \eq{eq:dm2fvmaster}, because as we now argue,
the ultraviolet divergences cancel in the difference of the sum and the
integral. First, if we replaced the momentum sum by the one, which corresponds
to $\qedTL$, the difference must be finite. This is because in $\qedTL$ only
the single $k=0$ mode is removed from the momentum sums, and these sums define
the integral in the $L,T\to\infty$ limit.  Furthermore the difference between
$\qedL$ and $\qedTL$ one loop self-energies is a one-dimensional integral over
$k_0$ with $\vec{k}=0$, which is ultraviolet convergent in the point-like case,
because of the $1/k_0^2$ suppression from the photon propagator. Since the form
factors, which arise in the vertex functions of composite particles are
expected to suppress ultraviolet modes, this integral will be ultraviolet
convergent also for composite particles. From this reasonable assumption it
follows that the difference of the one-loop self-energies in IV QED and
$\qedL$ in \eq{eq:dm2fvmaster} is free of ultraviolet divergences.

To obtain the desired asymptotic expansion of $\Delta m^2(T,L)$ in
powers of $1/L$, we apply the same Poisson formula as the one used in
\eq{eq:fvintegrals}. Power-law volume corrections
result from infrared singularities in the integrand. We
saw that $(p^2+m^2)D(p)$, $\Gamma_\mu(p,k)$ and
$\tilde\Gamma_{\nu\mu}(k';p,k)$ are analytic functions of their arguments in
the domain of integration of \eq{eq:dm2fvmaster}. Thus, the only
singularities that can arise are those corresponding to the poles of
the free photon propagator, $1/k^2$, and of the free scalar
propagator, $D_0(p+k)=1/(2k{\cdot} p+k^2)$, for $p^2=-m^2$. These
singularities show up in the limit $\vec{k}\to 0$. It should be noted
that there is an additional source of power-law corrections in
\eq{eq:dm2fvmaster}: the subtraction of the $\vec{k}=0$ contribution
in the sum over photon momenta.

Dimensional
arguments suggest that corrections up to and including $1/L^2$ can be
obtained by expanding the numerator of the
$\Gamma_\mu\cdot D\cdot\Gamma_\mu$ term in \eq{eq:dm2fvmaster} to
$\mathcal{O}(k_\mu)$ and that of the $\tilde\Gamma_{\mu\mu}$ term to
$\mathcal{O}(k_\mu^0)$. Setting $p^2=-m^2$, we perform this expansion
around the on-shell point $k^2=0$ and $(p+k)^2+m^2=P{\cdot} k=0$, with
$P\equiv 2p+k$, considering only terms up to the desired order in $k_\mu$
at the end of the calculation. This expansion can be performed because
these functions are analytic in the domain of integration of
\eq{eq:dm2fvmaster}. For the derivation, it is
actually convenient to push the expansion to one order higher in
$k_\mu$, as it allows to include the full point-particle
contribution. Note that we are expanding here the IV functions, and
can therefore use all of the properties of these functions in infinite volume,
including Lorentz covariance.

We begin with the propagator
\bea
D(p+k)&=&\frac1{P{\cdot} k}\left[1-(P{\cdot} k) \delta D(P{\cdot} k)\right]^{-1},
\eea
where $\delta D(P{\cdot} k)$ is the full self-energy of the composite scalar at $e=0$. It can be expanded as
\bea
\labell{eq:propexp}
D(p+k)= \frac1{P{\cdot} k} + \delta D(0)+(P{\cdot} k)\{\delta D(0)^2+\delta D'(0)\}+\ord{(P{\cdot} k)^2}\ . 
\eea
The two vertex functions are then decomposed into linear combinations
of invariant functions using Lorentz covariance and the properties of
these functions under discrete spacetime symmetry
transformations. 
Substituting the invariant decompositions into the WTIs of \eqs{eq:3ptWTI}{eq:4ptWTI}, 
expanding the resulting equations to the desired order in $k_\mu$ and
matching order by order yields:
\bea
\Gamma_\mu(p,k)/q &=& P_\mu\left[F(k^2)-\delta D(0)(P{\cdot} k)-
\delta D'(0)(P{\cdot} k)^2\right] - k_\mu F'(0)(P{\cdot} k) +\mathcal{O}(k^3)\ ,
\labell{eq:3ptexp}\\
\tilde\Gamma_{\nu\mu}(k;p,k)/q^2&=&-2\delta_{\mu\nu}+ 8 p_\mu p_\nu \delta D(0)+\mathcal{O}(k)\labell{eq:4ptexp}
\ ,
\eea
where $F(k^2)$ is the on-shell electromagnetic form-factor of the scalar. Its first derivative gives
the electromagnetic radius $\langle r^2\rangle= -6F'(0)$.
Then, replacing the propagator and vertex functions
in \eq{eq:dm2fvmaster} by the above expansions,
one finds that the terms
in $\delta D(0)$ and $\delta D'(0)$ cancel and one obtains
the point-particle result of \eq{eq:Dmuniversal} plus a finite term proportional to
$\langle r^2\rangle/L^3$
and a residual
\be
\Delta m^2_\mathrm{res}(T,L)= -\left[\sumint_k^\prime-\int\frac{d^4k}{(2\pi)^4}\right]\frac{1}{m^2k^2}\left\{\frac{k^2(p{\cdot}k) f_1+(p{\cdot}k)^3 f_2/m^2}{2k{\cdot} p+k^2}+k^2f_3+(p.k)^2f_4/m^2\right\}
\ .
\labell{eq:compscalFVres}
\ee
In this expression, the dimensionless, nonperturbative functions
$f_i=f_i(p{\cdot}k,k^2)$, $i=1,\cdots,4$ must fall off fast enough as
$|k_\mu|\to\infty$ for this residual to be finite. Indeed, as
discussed above, $\Delta m^2(T,L)$ is finite. Moreover, the terms
already computed in the $1/L$ expansion are also finite and by
definition these functions are such that
$f_i(p{\cdot}k,k^2)\underset{k_\mu\to 0}{\longrightarrow}
\mathrm{constant}$. Note that we have used crossing symmetry to
guarantee that contributions from $\tilde\Gamma_{\mu\mu}(k;p,k)$ in the
brackets begin at $\mathcal{O}(k^2)$.

Now, performing the change of variables $\vec{k}=2\pi\vec{n}/L$,
$\vec{n}\in \Z^3$, and taking the IV limit, we find:
\be \Delta m^2_\mathrm{res}(T,L)\underset{T,L\to+\infty}{\sim}
\frac{1}{m^2L^3} \left[\int
\frac{dk_0}{2\pi}\left\{\frac{m\,g_1(k_0)}{k_0+2im}+g_2(k_0)\right\}\right]\times
\left[\sum_{\vec{n}\in\Z^{3*}}-\int d^3n\right]1 \ , \ee
where the dimensionless, nonperturbative functions $g_i(k_0)$,
$i=1,2$, are derived from the $f_j(p{\cdot}k,k^2)$, $j=1,\cdots,4$,
with $\vec{k}=0$. These functions have a finite limit as $k_0\to 0$
and they fall off fast enough at large $k_0$ to guarantee that the
integral over $k_0$ is a finite number. Moreover, the difference
between the sum and integral of 1 over $\vec{n}$ is ${-}1$. Therefore
$\Delta m^2_\mathrm{res}$ is $\ord{1/L^3}$.

The spinor case proceeds in an entirely analogous way. The end result
is that the first two FV corrections to the mass of a composite
particle are identical for spin-0 and $1/2$ particles, and are equal
to those of point particles, as given in \eq{eq:Dmuniversal}. We use
this result throughout our numerical analysis of mass splittings.

In addition, in performing the expansions in
\eqs{eq:3ptexp}{eq:4ptexp}, it is clear that the next terms leading to
corrections of order $1/L^3$ or higher depend on the structure and
spin of the particles. This confirms the picture obtained in
\cite{Davoudi:2014qua}.

\subsection{Comparison to numerical QCD+QED results}

Here we compare the finite-volume behavior of hadron masses to the analytical
findings of the previous subsections. 

Let us begin with the kaon mass, whose volume dependence is shown in
Fig. 1 of the main text.  Panel A shows the neutral kaon
mass as a function of the box size. For the plot we used four ensembles
which differ only in box size, all other lattice
parameters being the same (see \sec{se:ens}).  There is no significant volume dependence in
accordance with \eq{eq:Dmuniversal}, since $q=0$.  In contrast, the charged
kaon mass is strongly dependent on the volume as can be seen in panel B
which displays the mass squared difference of the two particles. The
four different volumes cannot be described solely by the point-particle
behavior, which includes up to $1/L^2$ terms. This is a consequence of the
compositeness of the kaon and is nicely shown by our data.  If we add an
$1/L^3$ term to the formula, a nice fit quality can be achieved, shown as the
solid line on the plot. The $1/L$ and $1/L^2$ terms of the fit correspond to
dashed and dotted lines.

\begin{figure}[H]
\centering
\includegraphics{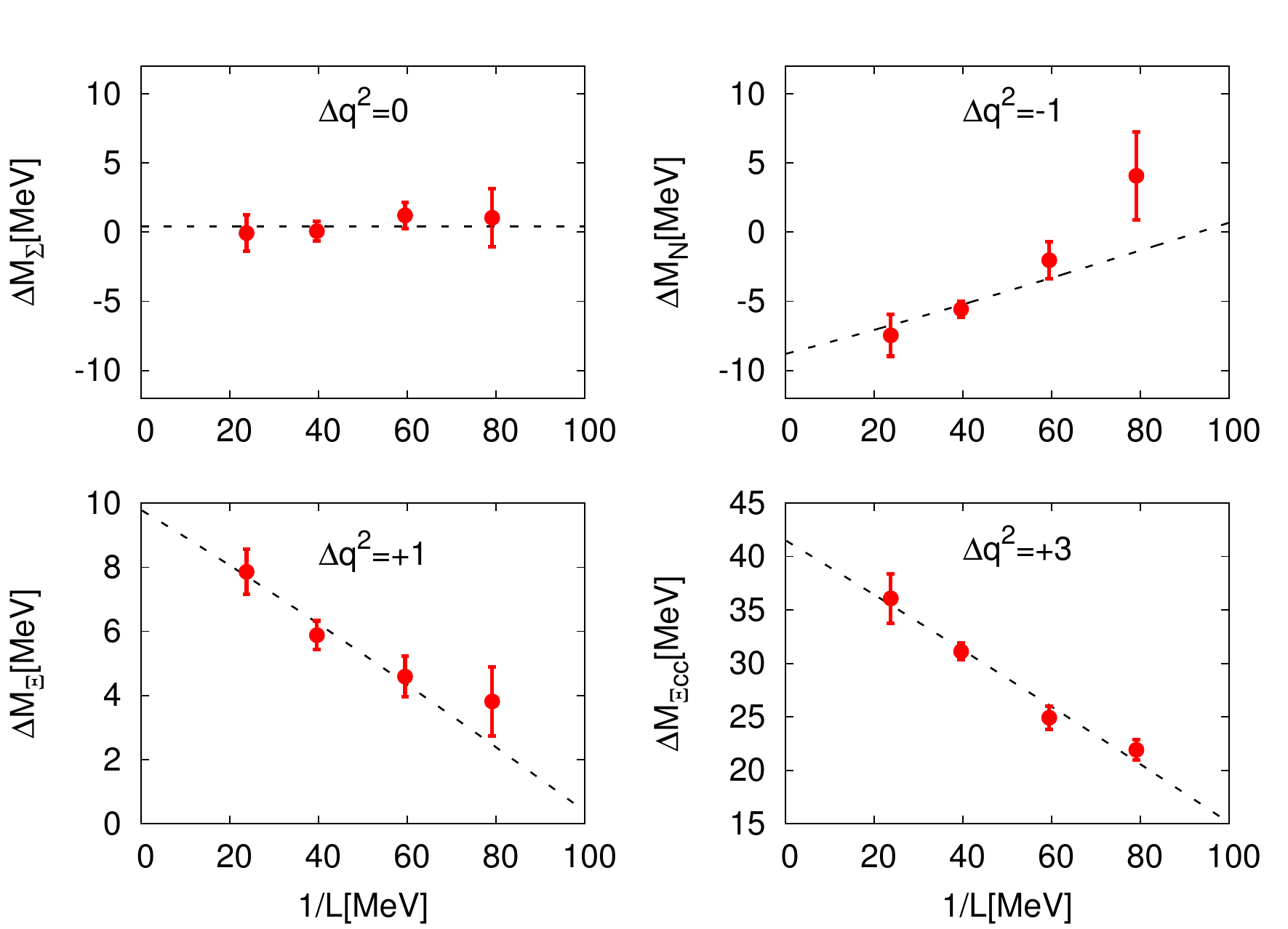}
\caption{\label{fi:fvol}
Finite-volume effects in baryon isospin splittings. The dependence
is always consistent with the universal behavior of \eq{eq:Dmuniversal} (dashed lines).}
\end{figure}

\fig{fi:fvol} shows the finite-volume dependence of various baryon mass
splittings. The precision in these channels is considerably smaller than for the
case of the kaon. All of them can be nicely described by the universal
behavior of \eq{eq:Dmuniversal} and the higher order term $1/L^3$ is always insignificant. The first
panel is the splitting of the $\Sigma$ baryons, $\Delta M_\Sigma=
M_{\Sigma-}-M_{\Sigma+}$. There is no significant finite-volume dependence at
all, since the absolute value of the two particles' charges are the same.
The other panels show splittings, where the charged squared difference is
non-zero. The dashed curves correspond to the universal fits. Since the $mL$ combination
is rather large for baryons, even the $\mathcal{O}(1/L^2)$ part is negligibly small.

\section{Algorithms}
\label{se:alg}

In this Section we describe the implementation of our updating algorithm. The
main novelty is the acceleration of the HMC by means of Fourier transformation,
which is essential to obtain a reliable update of the photon field. Another
algorithmic feature, which is taken from the literature, is the combined use
of multi-grid techniques and all-mode averaging (this will be discussed in
\sec{se:mass}).

\subsection{Hybrid Monte Carlo}

We use standard techniques to generate gauge configurations, including the
hybrid Monte-Carlo (HMC) algorithm \cite{Duane:1987de} with even-odd
preconditioning \cite{DeGrand:1990dk}, multiple time-scale integration
aka. ``Sexton-Weingarten scheme'' \cite{Sexton:1992nu}, mass preconditioning
aka. ``Hasenbusch trick'' \cite{Hasenbusch:2001ne}, Omelyan integrator
\cite{Takaishi:2005tz}, Rational Hybrid Monte Carlo acceleration with multiple
pseudofermions \cite{Clark:2006fx}. Configurations are separated by ten 
unit-length trajectories.

In previous isospin symmetric simulations, the two lightest quarks could be
bundled together and a standard Hybrid Monte Carlo was applicable.  Now that
all the quarks have different charges and masses, each of them has to be
included with a separate fermion determinant. This causes an extra difficulty,
because the sign of $\det D$ can become negative for small quark
masses when Wilson fermions are used. If this happens, the common update algorithms break down. 
However if the spectral gap of the hermitian $\gamma_5 D$ operator is large
enough, then the negative sign configurations play no role in the simulation.
This is the case, as one approaches the continuum or the infinite volume
limits \cite{DelDebbio:2005qa}.  In practice, if the lowest eigenvalue of the operator $\sqrt{D^\dagger D}$
has a non-vanishing separation from zero, then the number of 
negative-weight configurations is exponentially small in that separation and the standard
algorithms can be safely used.  If there are no negative sign configurations,
then the relation \[\det D= \det \sqrt{D^\dagger D}\] holds and the
simulation can be straightforwardly done with the RHMC algorithm. The RHMC also
needs the lowest eigenvalue of the $\sqrt{D^\dagger D}$ operator to be
determined, in order to get a suitable rational approximation for the (inverse) 
square root function.  The small eigenvalues are obtained using the Krylov-Schur
algorithm. Our implementation is based on the program package, which is
described in Ref.\ \cite{Hernandez:2005:SSF}.  In order to split the
determinant of the fermion matrix $D$ into an infrared and ultraviolet part, we
combine the RHMC algorithm with the Hasenbusch trick: \[ \det D = \det
\frac{1}{\sqrt{1+\rho^2/(D^\dagger D)}} \cdot \det \sqrt{ D^\dagger D+\rho^2},
\] and integrate the first/second determinant on a coarse/fine timescale. A
similar combination of the two techniques was already proposed in the staggered
fermion formulation \cite{Clark:2006wp}. 

\begin{figure}[H]
\centering
\includegraphics{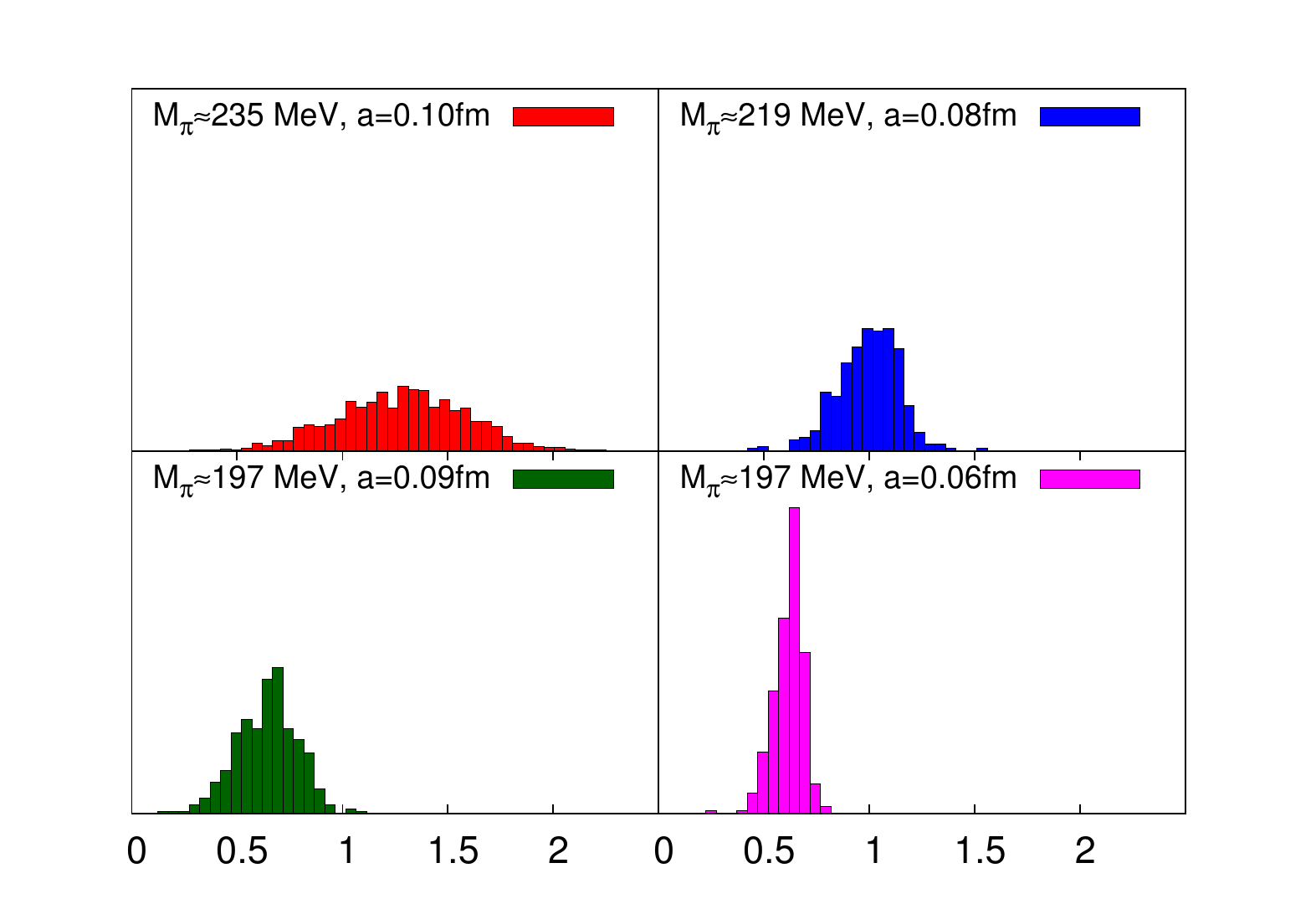}
\caption{Lowest eigenvalue of the $\sqrt{D^\dagger D}$ operator on the ensembles with
the smallest pion mass at each lattice spacing demonstrating the algorithmic stability.
\label{fi:lmin}
}
\end{figure}

The absence of the negative-weight configurations and the validity of the
absolute-value approximation can only be justified {\em a posteriori}.  This validation is shown
in Figure \ref{fi:lmin}.  The eigenvalue distributions are sufficiently bounded
away from zero, from which we conclude that our simulations are safe from
problems induced by eigenmodes becoming too small or even changing signs during
the RHMC integration, for the parameter values considered in this work.  We
also monitored the fermion force for dangerous spikes, which would signal
eigenmodes approaching zero and being repelled by a large force.  We do not see
any such spikes in the time history even in our potentially most affected
ensemble.

We check that the violation of the HMC energy ($\mathcal{H}$) conservation is
small and that the condition $\langle\mathrm{exp}(-\Delta
\mathcal{H})\rangle=1$ is always true within errors. 

Another important aspect that requires special attention is autocorrelations of
physical observables. Since we generate a Markov Chain with our HMC simulation
algorithm, subsequent measurements are correlated.  The most strongly
autocorrelated quantity is typically the topological charge. Its autocorrelation becomes
very large for lattice spacings much smaller than $\approx 0.05$~fm and our
choice of boundary conditions.  We calculated the autocorrelation of the
topological charge using the openly available package
UWerr~\cite{Wolff:2003sm}. To define the charge the Wilson-flow was used to
smoothen the configuration.  Even on the potentially most affected ensemble
with smallest lattice spacings the integrated
autocorrelation time was about 50 trajectories.

\subsection{Updating the photon field}

The HMC has to generate photon fields that fulfill the zero mode constraint,
\eq{eq:zm2}. Since the constraint is linear in the fields, the implementation
of such HMC is straightforward: one has to require that the HMC momenta also
satisfy the same constraint. This has to be enforced by removing the zero-mode
part of the usual HMC force in each microcanonical step.

\begin{figure}[H]
\centering
\includegraphics{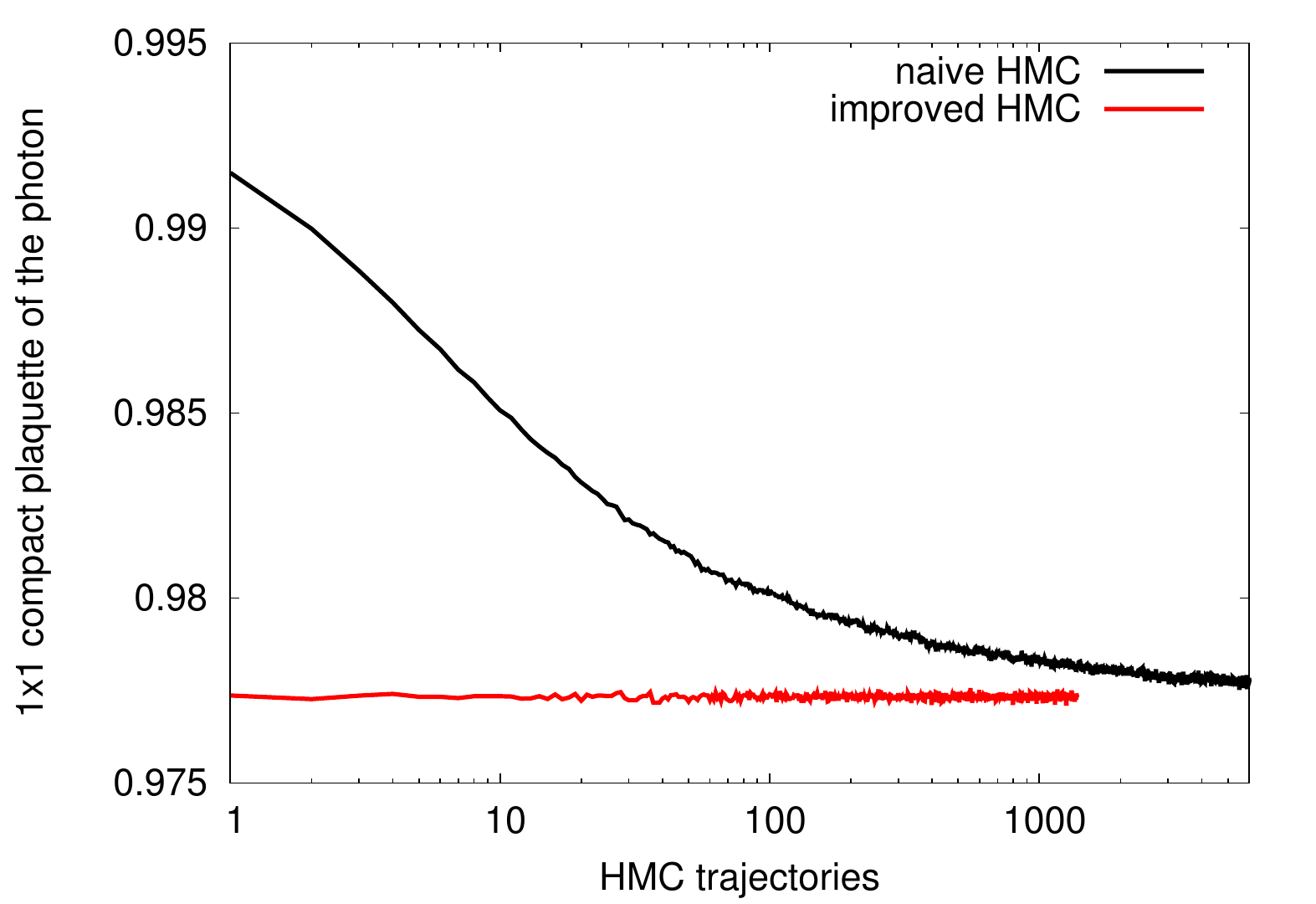}
\caption{\label{fi:alg}
Thermalization of the compact plaquette in the pure photon theory using the standard HMC and our improved variant.
}
\end{figure}

Due to the vanishing mass of the photon and the correspondingly large correlation
lengths a standard HMC integration of the photon fields results in large
autocorrelation times. Even for a pure photon simulation we observed that the
1x1 compact plaquette does not thermalize after several thousand trajectories,
see \fig{fi:alg}.  To eliminate this we use a Fourier accelerated HMC.
Let us first discuss the case of the pure photon theory. 
In the standard HMC we introduce momenta $\Pi_{\mu,k}$ for all
$A_{\mu,k}$'s and add a kinetic term $|\Pi_{\mu,k}|^2/(2m)$ to the photon action, \eq{eq:sgamma}.  The mass parameter is usually
chosen as $m=1$, but one can also allow it to be $k$-dependent, leading to the
following ''Hamiltonian'':
\[
\mathcal{H}=\frac{1}{2TL^3}\sum_{\mu,k} \left\{ |\hat{k}|^2 |A_{\mu,k}|^2+\frac{|\Pi_{\mu,k}|^2}{m_k} \right\}.
\]
At the beginning of an HMC trajectory we generate new momenta with a
distribution
\[\exp(-|\Pi_{\mu,k}|^2/(2m_k))\] and then integrate the canonical 
equations obtained from the above Hamiltonian until time $t$. The Hamiltonian is just a set of uncoupled
harmonic oscillators, therefore after time $t$ the photon field will be
\[
A_{\mu,k}(t)=A_{\mu,k}(0)\cos(\omega_k t) + \frac{\Pi_{\mu,k}(0)}{m_k\omega_k}\sin(\omega_k t)
\]
with $\omega_k=|\hat{k}|/\sqrt{m_k}$. For any oscillator if we let it evolve
for a quarter period ($\omega_k t=\pi/2$) then the first term becomes zero, so
the $A_{\mu,k}$ field ''forgets'' its previous ($t=0$) value. Since we have the
freedom to choose the parameters $m_k$ arbitrarily, setting $m_k=4|\hat{k}|^2/\pi^2$
will eliminate the first term for all oscillators at $t=1$, therefore
producing an exact heatbath step with zero autocorrelation. This is
demonstrated in \fig{fi:alg}, where the plaquette has the correct
distribution starting with the first trajectory.
The implementation of this algorithm requires Fourier transformations of the
photon fields and momenta for every step of the HMC, but this turned out to be
computationally negligible compared to Dirac operator inversions.

After including the coupling to quarks, the autocorrelation does not vanish
any more, but since the coupling is small it is much smaller than with a
standard HMC algorithm. In our runs the thermalization of the photon field always occurred
after a few trajectories.

\subsection{Cost issues}

Compared to a calculation with physical quark masses in the degenerate $2$
light flavor setup
the 1+1 light flavor framework doubles the computational costs.  Using two
different electromagnetic couplings increases the cost by another factor of
two.  Fixing the up-down quark mass difference to near its physical value
reduces the mass of the up quark by up to 40\%, which slows down the
inversions. The small mass region also requires large volumes to avoid
deceleration due to the fluctuations of the smallest Dirac eigenvalue toward
the origin.  To guarantee that the mean of the distribution of this eigenvalue
remains as many standard deviations away from zero as in our earlier studies,
while decreasing the mass by 40\%, requires increasing the volume by a factor
of 2.8. About a factor of five larger statistics are needed to have the desired
significance on the small neutron-proton mass difference.  In addition a more
costly analysis is required (the inversions were about twice as expensive as
the configuration generation, which gives a factor of three).  All in all one
has an increase in numerical cost by a factor of
$2\times2\times1.7\times2.8\times5\times3\sim300$. This increase by two to
three orders of magnitude is the reason why we did not attempt to extend our
simulations all the way down to the physical value of the pion mass as we did
in \cite{Durr:2010aw}. However, as shown in \cite{Durr:2008zz} and confirmed
here, simulations with pion masses down to $195\;\mathrm{MeV}$, as reached in
the present work, are sufficient to reliably determine hadron masses and
isospin splittings.

\section{Ensembles}
\label{se:ens}

We generated gauge ensembles at several different lattice parameter values
to be able to extrapolate to the physical quark mass point. The
following strategy was applied to find parameter values in the desired range, following
closely the method of Ref.\ \cite{Bietenholz:2010jr}.
First we
tuned, in $n_f=3+1$ simulations, the common light quark ($q$) mass and the
charm quark mass, so that the
pseudo-scalar meson masses approximately took the
following values:
\begin{align*}
M_{\bar{q}q}= 410\,\mev \quad\quad \text{and}\quad\quad
M_{\bar{c}c}= 2980\,\mev.
\end{align*}
The value for $M_{\bar{q}q}$ is approximately the average of the physical masses of the
mesons in the meson-octet with non-vanishing third component of isospin.
For the parameter tuning, we set the lattice spacing using the recently proposed
$w_0$ scale \cite{Borsanyi:2012zs}, which is based on the Wilson-flow \cite{Luscher:2010iy}.
The latter can be
measured easily and with high precision even on a handful of configurations.
In the exploratory runs, we took the value $w_0=0.1755$ fm, which was obtained using pure QCD simulations \cite{Borsanyi:2012zs}.
For the final results, we used the physical $\Omega$ mass for setting
the scale. In addition, we have redetermined the value of $w_0$ in our QCD+QED setup.
Afterwards we applied the method of Ref.\
\cite{Bietenholz:2010jr}: in a series of $n_f=1+1+1+1$ simulations we decreased
the masses of the up and down quarks and simultaneously increased the mass of
the strange quark, while keeping the sum of the three masses constant,
introducing a small splitting in the mass of the up and down quarks.
This line points towards the physical point with a very good precision. 
In this way we obtained 27 ensembles with no electromagnetic interaction, $e=0$. These are
the {\bf ``neutral ensembles''}.

At the bare parameters of some of these ``neutral ensembles'' we switched on
the electromagnetic interaction. Since this step introduces an extra additive
renormalization, the spectrum always changed significantly. We tuned the masses
by $\delta_u$, $\delta_d$ and $\delta_s$ additive shifts, so that the connected
meson masses for each quark took approximately the same values as in the
starting ``neutral ensemble'' \cite{Borsanyi:2013lga,Portelli:2010yn}.  In this
way we obtained our 14 {\bf ``charged ensembles''}. Four of these ensembles
have four different volumes, while all the other parameters are the same. They
were used to study the finite size effects (see \sec{se:fvol}). Similarly
there are four ensembles that have four different $e$ values and approximately equal pseuodoscalar masses. They can be used to test the
$e$-dependence of the results (see \sec{se:qflow}).

Altogether we have 41 ensembles at four different gluon gauge couplings.
\tab{ta:spa} lists the gauge couplings, the corresponding lattice spacings and
the bare charm mass. The lattice spacings are calculated using the
$\Omega$ mass as an input. The
detailed lists of our neutral and charged ensembles are given in
\tabs{ta:neu}{ta:cha}.

\section{Electric charge renormalization}
\label{se:qflow}

For small electromagnetic couplings there is a signal/noise problem.
Operators that are symmetric under charge conjugation, including the particle
propagators in this study, can only depend on $e^2$.
However on a given gauge configuration the operators in general have a linear
contribution in $e$, which vanishes after doing the ensemble average.  The
remaining signal is proportional to $e^2$, whereas the noise is of $O(e)$. This
linear noise term can be set to zero in the case of quenched QED: one averages the observables at couplings $+e$ and $-e$, canceling the linear term configuration by configuration \cite{Blum:2007cy}. However in the dynamical
case, this trick does not work, because the sign of the charge in the sea cannot be flipped while maintaining correlations in the gauge fields. Instead we decided to simulate at
couplings that are larger than the physical one, so that the signal
outweighs the noise. Then we interpolate the results, between those obtained with  $e>0$ and $e=0$, to the
physical value of the coupling. 

Here we have to discuss another issue: how to define the physical value of the
electromagnetic coupling? Ideally one would like to tune $e$ so that, in the
Thomson limit, i.e.\ at the scale of the electron mass, the renormalized
coupling $\alpha=e_R^2/(4\pi)$ takes on its experimental value, $\alpha^{-1}=137.036\dots$.
Evidently our lattices are not large enough to make measurements in this limit.
Instead we define the renormalized coupling at a hadronic scale. The difference
between the two is of order $\alpha^2$. If $\alpha$ in the Thomson-limit takes
the experimental value, the relative difference is on the percent level and can be
neglected in our results, since it is much smaller than other errors.

\begin{figure}[H]
\centering
\includegraphics{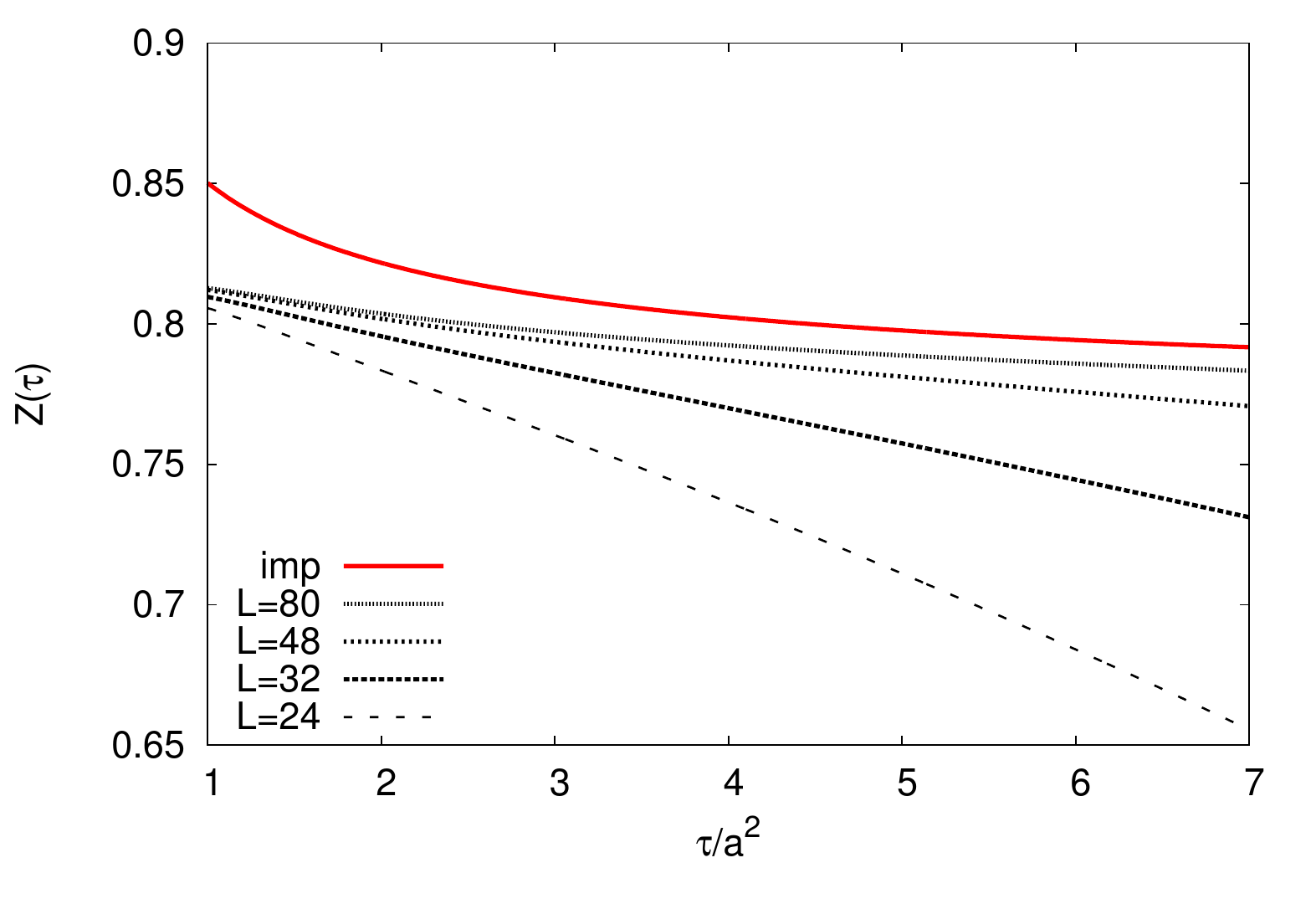}
\caption{\label{fi:qflowFV}
Renormalization factor of the electric charge as the function of flow time
on four ensembles with fixed bare parameters. The dashed
curves show $Z(\tau)$'s using $E_{\text{tree}}= 3/(32\pi^2)$ and have sizeable finite-volume dependence.
Using $E_{\text{tree}}(\tau)$ from \eq{eq:tree} eliminates volume dependence completely: the curves corresponding to the four volumes lie on top of one another (solid line).
}
\end{figure}

For the definition of a renormalized coupling we use the Wilson flow
\cite{Luscher:2010iy}. Now the photon fields are evolved with the gradient of
the photon action. Similarly to the gluonic case, we measure
along the flow:\[ E(\tau)= \frac{\tau^2 a^4}{TL^3}\sum_{\mu,\nu,x}
F_{\mu\nu,x}^{(A)}(\tau) F_{\mu\nu,x}^{(A)}(\tau),\] where
$F_{\mu\nu,x}^{(A)}(\tau)$ is the photon field-strength tensor at a Wilson-flow
time $\tau$.  Because $e^2E(\tau)$ does not renormalize\footnote{We use a
normalization of the photon field where $e$ is not in the photon action
but in the covariant derivative.} and is proportional to $e^2$, we can use it
to define the renormalized coupling at scale $\tau$: \[e_R^2(\tau)=
Z(\tau)\cdot e^2\quad\text{ with }\quad Z(\tau)= \langle E(\tau)\rangle/
E_{\text{tree}}(\tau)\] The proportionality constant requires the tree-level value $E_{\text{tree}}(\tau)$. In infinite volume and in the continuum limit, the tree-level value
does not depend on the flow time: it is $E_{\text{tree}}= 3/(32\pi^2)$,
see Ref.\ \cite{Luscher:2010we}. In finite volume and at finite lattice spacing with our definition
of $F_{\mu\nu,x}^{(A)}$ the
tree-level value becomes: \begin{align}\label{eq:tree} E_{\text{tree}}(\tau)= \frac{\tau^2}{T L^3}\sum_k
\frac{\exp( -2 |\hat{k}|^2\tau)}{a^2|\hat{k}|^2} \left[ \sum_{\mu\ne\nu} (1+\cos
ak_\nu)\sin^2 ak_\mu \right].  \end{align} \fig{fi:qflowFV} shows the renormalization constant
$Z(\tau)$ as a function of flow time for the runs of the finite-volume
scaling study.  We plot $Z(\tau)$ using  with the infinite-volume, continuum
tree-level value with dashed lines and with \eq{eq:tree}  with solid lines.
Carrying out this step for all four volumes ($L=24, \dots, 80$) the 
resulting lines lie on top of each other.
As one can see all volume dependence is removed by the improved tree-level formula.

The difference of the renormalized coupling at different flow times is an
$O(\alpha^2)$ effect, which at the physical value of $\alpha$ is negligible in
our calculation. However, in order to be conservative, we carry out the final
analysis using two different $\tau$'s, so that we can estimate systematic
effects arising from this choice.  Since in the fermion-photon coupling the
photon fields have undergone smearing, we also use the smeared photon field in
the definition of $E(\tau)$. This changes the renormalized coupling only by
lattice artefacts.

\begin{figure}[H]
\centering
\includegraphics{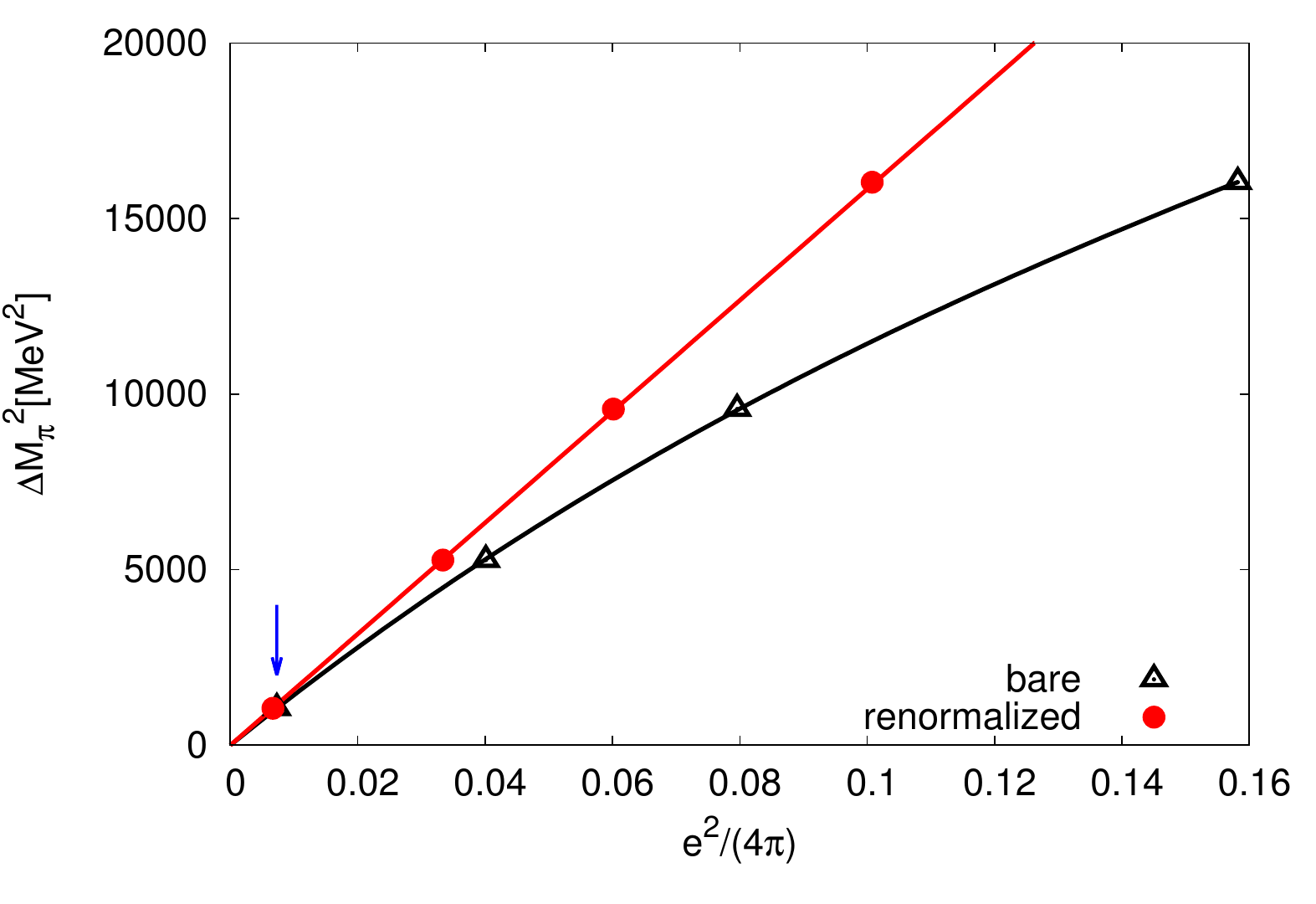}
\caption{\label{fi:qflow}
Pion splitting as a function of the bare and renormalized couplings
for four charged ensembles. The solid lines are linear/cubic fits to the renormalized/bare data points.
The arrow indicates the physical value of $\alpha$.
}
\end{figure}

The renormalized coupling defined above has an important benefit in combination
with simulating at larger than physical couplings. For the sake of the
illustration let us consider the mass difference \[\Delta M_\pi^2=
M_{\bar{d}u}^2-\frac{1}{2}\left( M_{\bar{u}u}^2+M_{\bar{d}d}^2\right),\] which
we call pion splitting.  Here $M_{\bar{q}q}$ are masses of neutral mesons,
whose propagator includes only quark-connected diagrams. Using partially
quenched chiral perturbation theory one can show that, at leading order in
isospin breaking, this quantity is purely electromagnetic: there is no
contribution from the quark mass difference \cite{Borsanyi:2013lga,Bijnens:2006mk}.
At the same time this is the isospin splitting most precisely obtained 
from our simulations and therefore
 any deviation from the linear behavior
in $\alpha$ can be observed.  \fig{fi:qflow} shows the data of the four charged ensembles, that
have four different $e$ values and approximately equal pseudoscalar masses.
The solid lines are polynomial fits to the data
without a constant term.  As a function of the bare coupling (open symbols) the
splitting has a significant curvature. If we plot it as a function of the
renormalized coupling (filled symbols), we see no deviation from linear
behavior. Here an illustrative scale was chosen at $(8\tau)^{-1/2}=400\,\mev$. This observation is true for other observables considered in this
paper: no significant $O(\alpha^2)$ contribution can be seen for the whole
coupling range, which is also true for the other scale choices used in the final analysis.

\section{Extracting hadron masses}
\label{se:mass}
To calculate hadron propagators we fix the smeared photon and gluon fields to
Coulomb gauge. We use local hadron operators, in which the quark fields are
Gaussian smeared, with a smearing radius of about $0.3$ fm.  We observe, 
as
in many other studies that reducing the noise in hadron propagators can
be efficiently done by using many different quark sources per gauge
configuration. In the case of the nucleon propagator we use several hundred
different source positions.  For these propagator calculations we use the
2-level multi-grid approach of Ref.\ \cite{Frommer:2013fsa} and also the
variance reduction technique of Ref.\ \cite{Blum:2012uh}.

We fold the propagators around the midpoint of the lattice in time 
and fit them with standard
one-state functions. To obtain the mass splittings our method is to jointly fit
the isospin partners with the mass difference, the average mass and the two
amplitudes as fit parameters~\cite{Borsanyi:2013lga}. 
Fitting the propagators separately and
subtracting the fitted values yields results that are consistent with the
joint fit method and the statistical errors are of similar size. All plateau
fits take into account the correlation between the timeslices and isospin partners. 
In
order to have a stable correlation matrix we only fit ten time slices at once.

Selecting the fit interval for a given hadronic channel is a highly non-trivial
task. In order to get rid of the excited states one wants to take large values
for the timeslice, where the mass fit starts ($t_{min}$). 
However, taking values that are too large
results in suboptimal signal/noise ratios. Since our goal is
to reach precisions on the sub-permil level, a reliable, robust and controlled
method is needed to choose the proper fit intervals.

\begin{figure}[H]
\centering
\includegraphics{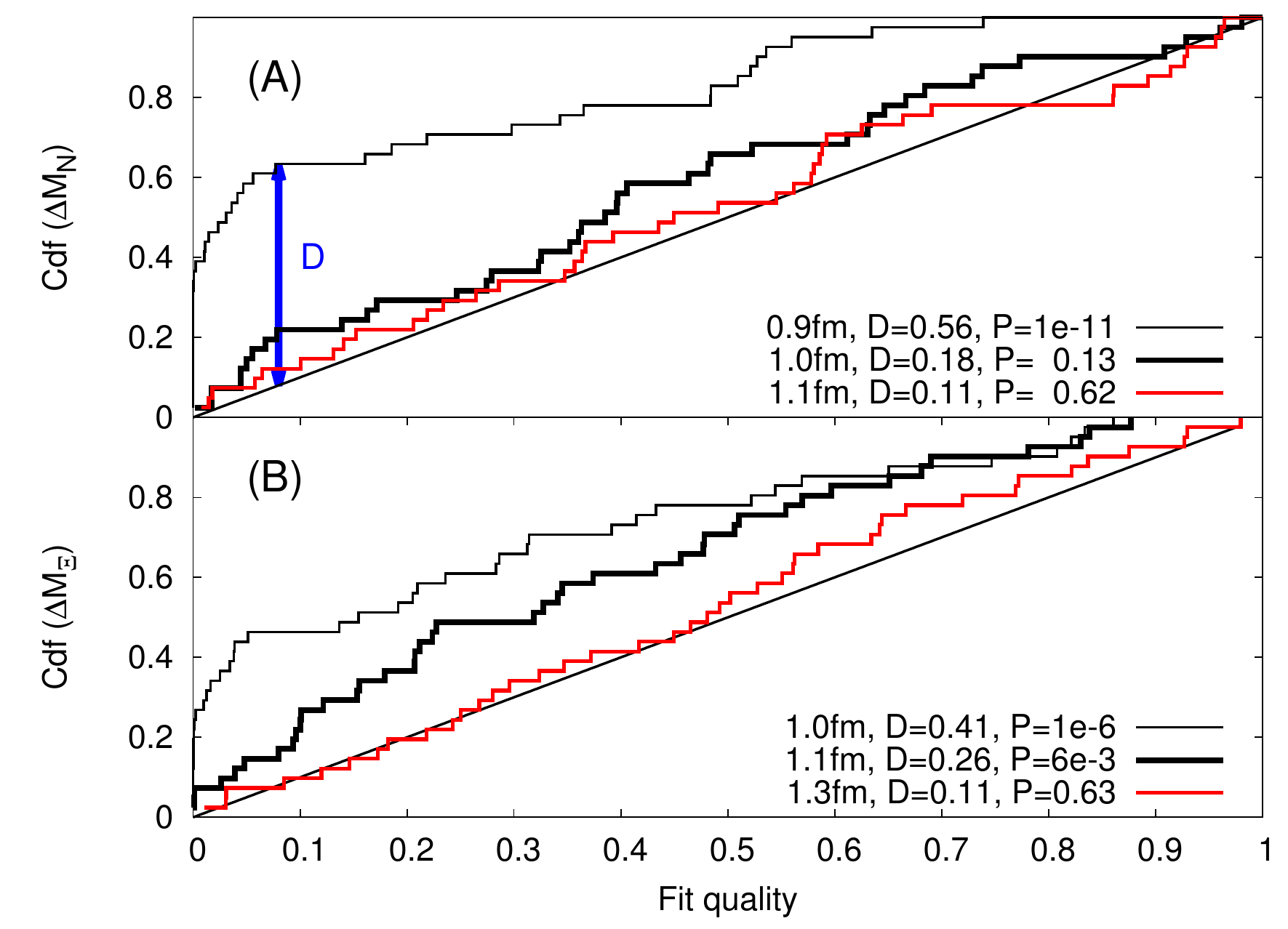}
\caption{\label{fi:distro}
Cumulative distribution functions of the fit qualities. The distribution
is obtained using the fits for the $\Delta M_N$ (panel A) and $\Delta M_\Xi$ (panel B) mass difference on the 41 ensembles. Three different
measured distributions are plotted using three different $t_{min}$ values, the largest ones correspond to our choices in the final analysis.
The straight line
corresponds to the expected uniform distribution.
An arrow shows
the maximum distance $D$ for one of the distributions. For each value of 
$t_{min}$ we provide the distances $D$ and the probabilities $P$ that the observed distributions are uniform.
}
\end{figure}

The most popular way to select a good fit range is to calculate the correlated
$\chi^2$ of a mass fit and use ranges for which the $\chi^2$/dof is about one.
Note that this procedure is not fully satisfactory. The values of the $\chi^2$/dof
for a given hadron channel should follow instead the $\chi^2$ distribution.
Thus we want to determine the probability that from a given $t_{min}$ the $\chi^2$/dof
values follow the expected distribution (or equivalently the corresponding
goodnesses of the fits are uniformly distributed). Since we have 41 independent ensembles
it is possible to look at the distributions of the $\chi^2$/dof for
various $t_{min}$ values. \fig{fi:distro} shows the expected cumulative
probability distributions of the fit qualities (which has to be linear) and the
measured distribution. Both the nucleon and $\Xi$ channels are depicted. As
expected the larger the $t_{min}$ the closer the measured distribution is to the
expected one.

The standard way to statistically interpret \fig{fi:distro} is to carry
out a Kolmogorov-Smirnov analysis \cite{Press:1992zz}. Thus one determines the
maximum value of the absolute difference between the expected and measured
cumulative probability distributions (D). This is then used to define a
significance level or probability (P) that the measured distribution can indeed
be one originating from the expected uniform distribution. We start our fits
from $t_{min}$ values, for which this probability is larger than 0.3. In
addition, to estimate the possible contribution of excited states, 
we use another fit range starting from one timeslice later. (Since
the excited states have even smaller contributions for fit ranges which start
even later, one might use them too. However, due to the increasing noise this
provides no new information). Obviously, for different hadronic channels
the proper $t_{min}$ values are different. They are collected in \tab{ta:tmin}.

\section{Reaching the physical point}
\label{se:iso}

Our present procedure closely follows our earlier
studies of the hadron spectrum \cite{Durr:2008zz}.
We consider two different paths, in bare parameter space, to the physical mass
point and continuum limit. 
These two techniques
correspond to two different ways of normalizing hadron masses and their isospin
splittings for a set of fixed parameters. For both methods we follow several
strategies for the extrapolation to the physical mass point and to the
continuum limit. The physical mass point is defined by
$M_{\pi^+}=139.570\,\mev$, $M_{K^+}=493.68\,\mev$, $M_{K^0}=497.61\,\mev$,
$M_{D^0}=1864.9\,\mev$ and the electromagnetic coupling in the
Thomson-limit $\alpha^{-1}=137.036$ \cite{Beringer:1900zz}. A particularly
important quantity of the analysis is the kaon mass splitting. Its experimental
value is well known and it can be measured with high accuracy on the lattice. We
use the value $\Delta M_K^2$=$M_{K^0}^2$-$M_{K^+}^2=3896\,\mev^2$. For both
methods the lattice spacing at a given $g^2$ was determined by the Omega baryon
mass at the physical mass point (its experimental value 
is $M_\Omega=1672.4\,\mev$). The values for $a$ are given in \tab{ta:spa}.

\subsection{Normalizing hadron masses}
\label{se:ratio}
We call the two methods of normalizing the hadron masses and their splittings:
a. ``mass-independent scale setting'' and b. ``ratio method''.

a. The mass-independent scale setting is the more traditional technique. One
takes the lattice spacing from \tab{ta:spa} and expresses all the hadron masses
and their isospin splittings in physical units using this $a$. Thus, the mass
values measured on the lattice are divided/normalized by the lattice spacing.
The splittings are then extrapolated to the physical mass point and continuum
limit as explained below.  Since this is the more traditional technique we
explain the extrapolations and interpolations using this normalization
prescription (for the ratio method the corresponding formulae can be obtained
straightforwardly).

b. The ratio method is motivated by the fact that in QCD+QED one can calculate
only dimensionless combinations of observables, e.g. mass ratios (or isospin
splittings normalized by some hadron mass). Furthermore, in such ratios
cancellations of statistical uncertainties and systematic effects may occur.
The method uses the mass ratios $M_{\pi^+}/M_\Omega$, $M_{K^+}/M_\Omega$,
$M_{K^0}/M_\Omega$, $M_{D^0}/M_\Omega$ as input parameters and expresses the splittings
normalized by $M_\Omega$.

\subsection{Determining the isospin splittings}
\label{se:isofun}
There are two sources of isospin violation: electromagnetism  and the mass
difference of the up and down quarks. The isospin splittings are expanded in
powers of the renormalized coupling $\alpha=e_R^2/(4\pi)$ and the quark mass
difference $\delta m=m_d-m_u$.  This expansion is expected to converge rapidly
at the physical values of these parameters.  
However we work at somewhat larger electromagnetic
coupling values. As was shown in \sec{se:qflow}, higher-order terms in
$\alpha$ are negligible if we use a coupling defined at a hadronic scale. So we
work only with the linear terms, i.e. in $\mathcal{O}(\delta m,\alpha)$.
In linear order of isospin breaking, an arbitrary mass splitting $\Delta
M_X$ can be written as a sum of two terms:
\begin{align}
\Delta M_X= F_X(M_{\pi^+},M_{K^0},M_{D^0},L,a) \cdot \alpha + G_X(M_{\pi^+},M_{K^0},M_{D^0},a) \cdot \Delta M_K^2,
\label{eq:dx}
\end{align}
where $F_X,G_X$ are functions of pseudoscalar meson masses and the lattice spacing.
The QED part of $\Delta M_K^2$ has an $L$ dependence but this can be
absorbed into the $L$ dependence of $F_X$. The charged particle
masses that enter in \eq{eq:dx}, are already corrected for the universal
finite-size effect \eq{eq:Dmuniversal}.  Higher order polynomial finite-size
effects, starting with $1/L^3$, are still allowed in the electromagnetic part
$F_X$. Since $\Delta M_K^2$ is the mass squared difference of a neutral and
charged particle, the second term of \eq{eq:dx} mixes the strong and
electromagnetic isospin breaking effects. This is no problem of principle and
good quality fits can be achieved in practice. An alternative
procedure, where $\Delta M_\Sigma$ was used instead of $\Delta M_K^2$
was also performed and yielded compatible results.

We choose several different functional forms for $F_X$ and $G_X$. The difference
in the results coming from different choices will be part of the systematic error (see
\sec{se:fit}). In all cases we have in $F_X$ and $G_X$ a constant term
and a pion-mass dependence parametrized by $M_\pi^2$. $F_X$ always contains a term
proportional to $1/L^3$ and $G_X$ a term parameterizing the lattice spacing
dependence, either with $g^2 a$ or $a^2$. 
Note that our action
has a leading cutoff dependence of $O(g^2a)$, but in order to
give a more conservative estimate of the pertaining systematic error
we added the alternative ansatz.
Optionally we add $M_{K^0}^2$, $M_{\pi^+}^4$
and $M_{D^0}$ dependencies. 
In many cases, especially for the neutron-proton mass difference,
the coefficients corresponding to all but the constant terms are consistent
with zero.

\subsection{Separating QED and QCD effects}
Here we show how to separate the electromagnetic and strong
contributions in an isospin splitting, i.e.\  
\[\Delta M_X= \Delta_\mathrm{QED}M_X +
\Delta_\mathrm{QCD}M_X,\]
where the first and second terms are proportional to
$\alpha$ and $\delta m$, respectively.  It is clear, that it is sufficient to decompose the kaon
mass squared difference, 
\[\Delta M_K^2= \Delta_\mathrm{QED}M_K^2 + \Delta_\mathrm{QCD}
M_K^2,\]
since the separation for the other splittings can be obtained from this
decomposition through \eq{eq:dx}.  Any such decomposition has an ambiguity of
$\ord{\alpha\delta m}$, which is NLO in isospin breaking, i.e.\ one order
higher than the one at which we work. This arises from the scheme dependence in
the definition of $\delta m$. In \cite{Borsanyi:2013lga} we proposed a
separation based on the difference of connected meson masses, $\Delta
M^2=M_{\bar dd}^2-M_{\bar uu}^2$. Using chiral perturbation theory
\cite{Bijnens:2006mk}, $\Delta M^2$ can be shown to be proportional to $\delta
m$ up to corrections of $\ord{\alpha^2,\alpha\delta m, (\delta m)^2, \alpha
m_{ud}}$, which are also NLO if we make the very reasonable assumption that
$\ord{m_{ud}}\sim\ord{\delta m}$. Though this technique is sufficient for the
electroquenched approximation used in previous work, it has a drawback here, in
that these connected mesons are not in the spectrum of the full theory. One
could use this approach in a partially quenched framework, but we want to avoid
this in the present {\em ab initio} setup.

Here we show that, within the precision of our calculation, an alternative
separation can be applied. The method is based only on observed particles.
Since the $\Sigma^+$ and $\Sigma^-$ have the same charge squared and the same
spin, their electromagnetic self-energies are identical at $\ord{\alpha}$, if
these particles are assumed to be point-like. In that case, $\Delta
M_\Sigma=M_{\Sigma^-}-M_{\Sigma^+}$ comes purely from $\delta m$. Because these
particles are composite, there are corrections to this result.  We now argue
that they are no larger than the $0.2\,\mev$ overall uncertainty on our
determination of $\Delta M_\Sigma$.  In the seminal quark mass review
\cite{Gasser:1982ap}, Gasser and Leutwyler find $\Delta M_\Sigma
=0.17(30)\,\mev$ by evaluating the Cottingham formula (for a recent discussion see \cite{WalkerLoud:2012bg}). Using $\Delta M^2=0$ to
define the electromagnetic contribution to splittings
we found $\Delta_\mathrm{QED} M_\Sigma =
-0.08(12)(34)\,\mev$ in our electroquenched lattice computation \cite{Borsanyi:2013lga}, which is
consistent with 0, albeit within relatively large uncertainties. Repeating this
analysis in the fully unquenched theory we find
$\Delta_\mathrm{QED}M_\Sigma = 0.18(12)(6)\,\mev$, again consistent with 0
within a little more than one standard deviation.  All of these determinations
provide quantitative evidence that our proposed QCD-QED separation convention,
i.e.\ using \[\Delta_\mathrm{QED}M_ \Sigma = 0,\] is adequate within the
precision of the present paper.
This choice establishes a first benchmark against which future results can be compared. It can certainly be refined when calculations, such as the one presented here, become more precise.

Applying the fit formula \eq{eq:dx} to the $\Delta M_\Sigma$ splitting and carrying
out the fit procedure as described in the next section we obtain
\begin{gather}
\label{eq:dqedmka}
\Delta_{QED} M_K^2= -2250(80)(90)\text{ MeV}^2,
\end{gather}
where the first error is statistical, the second is the systematic error. The separation
for the other isospin splittings is based on this finding. The results are given in Table 1 in the main text.

\subsection{The Coleman-Glashow relation}
In addition to the mass splittings, we also compute
$\Delta_\mathrm{CG}=\Delta M_N-\Delta M_\Sigma+\Delta M_\Xi$, which parametrizes
the violation of the Coleman-Glashow relation, 
$\Delta_\mathrm{CG} = 0$~\cite{Coleman:1961jn,Zweig:1964jf}. From quark exchange
symmetries we know that at vanishing $\alpha$ the leading contribution to
$\Delta_\mathrm{CG}\propto (m_s-m_d)(m_s-m_u)(m_d-m_u)$, see e.g. Ref.\ \cite{Horsley:2012fw}. At non-vanishing $\alpha$
there is a remnant $d\leftrightarrow s$ exchange
symmetry yielding the leading order contributions
\begin{equation}
\Delta_\mathrm{CG}=\hat{F}_\mathrm{CG}\cdot\alpha \cdot(m_s-m_d) +\hat{G}_\mathrm{CG}\cdot(m_s-m_d)(m_s-m_u)(m_d-m_u).
\end{equation}
We now trade the quark mass differences for mass-squared differences of suitable pseudoscalars. We define
$M_{sd}^2=(M_{K+}^2-M_{\pi+}^2)/2$, which to leading order is
proportional to $(m_s-m_d)$. For a quantity, that is proportional to $(m_d-m_u)$ 
we take $M_{du}^2= \Delta_{QCD}M_K^2$, where we use the convention for the separation of QCD and QED effects
from the previous subsection.
Finally we define $M_{su}^2=M_{sd}^2+M_{du}^2$
and fit
\begin{equation}
    \Delta_\mathrm{CG}=F_\mathrm{CG}(L,a)\cdot \alpha \cdot M_{sd}^2+ G_\mathrm{CG}(a)\cdot M_{sd}^2M_{su}^2M_{du}^2,
\label{eq:fcg}
\end{equation}
where we can include discretization terms $g^2 a$ or $a^2$ and finite-volume terms into
the fit functions $F_\mathrm{CG}$ and $G_\mathrm{CG}$.
A fit of our results to \eq{eq:fcg} 
yields $\Delta_\mathrm{CG}=0.00(11)(06)\,\mev$: 
the Coleman-Glashow relation is satisfied to high accuracy.

\section{Akaike's information criterion}
\label{se:aic}

When we have several candidate models to fit our lattice results, e.g.\ several
fit functions with a different number of parameters, or several fit ranges in
Euclidean propagator fits, Akaike's information criterion (AIC) gives a
guideline on how to weigh the results of the various models.

Let $\Gamma_1, \dots, \Gamma_n$ denote the result of $n$ independent
measurements from the same (unknown) probability 
distribution $g^{(1)}(\Gamma)$,
e.g.\ of the Euclidean propagator of a hadron measured on $n$ independent
configurations. Let $f^{(1)}(\Gamma| \theta)$ be the model which is used to
approximate the true distribution $g^{(1)}(\Gamma)$, depending on $p$ parameters,
the parameter vector being denoted by $\theta$.

Akaike proposed \cite{Akaike1973} to measure the distance of the model $f$ from
the true distribution $g$ using the Kullback--Leibler divergence
\cite{Kullback1951}
\begin{align}
    I(g,f(\theta))= 
    \int \mathrm{d}\Gamma \ g(\Gamma) \log \left( \frac{ g(\Gamma) }{ f(\Gamma | \theta) } \right)
    = \int \mathrm{d}\Gamma \left\{ g(\Gamma) \log \left[ g(\Gamma) \right] - g(\Gamma) \log\left[ f(\Gamma | \theta) \right]\right\},
\label{eq:aic_KL}
\end{align}
where $g(\Gamma) = \prod_{k=1}^n g^{(1)}(\Gamma_k)$ and $f(\Gamma | \theta) =
\prod_{k=1}^n f^{(1)}(\Gamma_k | \theta) $ denote the joint distributions of the $n$
trials and $\mathrm{d}\Gamma= \prod_{k=1}^n \mathrm{d}\Gamma_k$ is the joint integration
measure. When the $n$ independent samples $\Gamma_1,\dots,\Gamma_n$ are drawn
from the model probability distribution $f(\Gamma|\theta)$, then in the $n
\to\infty$ limit, the negative of this divergence
gives the logarithm of the probability that the sample becomes distributed
according to the distribution $g(\Gamma)$ \cite{Sanov1961}. Therefore,
when we have several different models, $f_m(\Gamma | \theta_m)$,
where $m$ is the index and $\theta_m$ are the parameters of the given model, then
the weight
\begin{equation}
    w_m = \frac{\exp\left[-I(g,f_m(\theta_m))\right]}{\sum_{m'}
    \exp\left[-I(g,f_{m'}(\theta_{m'}))\right]}
\label{eq:aic_wiB}
\end{equation}
for model $m$ emerges naturally.  That is, we are considering each model with
a weight proportional to the probability that the given model reproduces the
measured data.

Once we have the outcome $\Gamma$ of a series of $n$ independent measurements,
in order to test the data against our models, we need to perform two
steps. First, for each model $m$ we need to find the parameter
$\hat{\theta}_{m,\Gamma}$ that minimizes the function $I(g,f_{m})$ in
\eq{eq:aic_KL}. In the large $n$ limit, under mild regularity conditions,
the maximum likelihood estimate minimizes \eq{eq:aic_KL} with
probability one \cite{Huber1967}, therefore we take $\hat{\theta}_{m,\Gamma}$
to be the maximum likelihood estimate.

The second step is to estimate
$I(g,f_{m}(\hat{\theta}_{m,\Gamma}))$. The first term of
\eq{eq:aic_KL} does not depend on the model, therefore it cancels
out when the weights in \eq{eq:aic_wiB} are calculated. Thus, the quantity to
be estimated is
\begin{equation}
J_{m}(\Gamma) = \int \mathrm{d}\Gamma'\ 
g(\Gamma') \, \log \left[ f(\Gamma' |
  \hat{\theta}_{m,\Gamma})\right].
\label{eq:aic_T}
\end{equation}
The integral cannot be directly performed, since the true probability
distribution $g$ is unknown. Nevertheless, Akaike \cite{Akaike1973} suggests to
estimate \eq{eq:aic_T} with
\begin{equation}
-\frac{1}{2} \, \text{AIC}_{m}(\Gamma) = \log \left[ f(\Gamma |
\hat{\theta}_{m,\Gamma})\right] - p_{m},
\label{eq:aic_AICdef}
\end{equation}
where $p_{m}$ is the number of parameters in model $m$. If the model
distribution is close to the true distribution, then in the large $n$ limit
\begin{equation}
    \int \mathrm{d}\Gamma\ g(\Gamma) \left[ -\frac{1}{2} \, \text{AIC}_{m}(\Gamma) \right] \,
    \approx \int \mathrm{d}\Gamma\ g(\Gamma) \left[ J_{m}(\Gamma) \right],
\label{eq:aic_AICexp}
\end{equation}
that is, in expectation \eq{eq:aic_T} and \eq{eq:aic_AICdef} agree.
Therefore, the estimation of $J_{m}(\Gamma)$ with $-\frac{1}{2} \,
\text{AIC}_{m}(\Gamma)$ is justified. For a concise derivation of
\eq{eq:aic_AICexp} see Ref.\ \cite{Kitagawa1996}, or see
Ref.\ \cite{Konishi2008} for a more detailed exposition.
Finally, the results obtained from the various models have to be weighed using
the Akaike weights \cite{Akaike1978c}
\begin{equation}
    w_{m} = \frac{\exp\left(-\frac{1}{2} \, \text{AIC}_{m}(\Gamma) \right)}{\sum_{m'}
  \exp\left( -\frac{1}{2} \, \text{AIC}_{m'}(\Gamma) \right)}.
\label{eq:aic_AICw}
\end{equation}

For models with normally distributed errors, Akaike's information criterion
takes the simple form
\begin{equation}
\text{AIC}_{m}(\Gamma) = \chi^2_m + 2p_m,
\label{eq:aic_aic_chi2}
\end{equation}
where all additive constants independent of the models were dropped.
That is, the AIC weight prefers the models with lower $\chi^2$ values, but
punishes the ones with too many fit parameters.

\section{Final results and systematic errors}
\label{se:fit}

Hadron masses or mass splittings in the continuum limit at the physical point
are extracted from lattice data following a two step procedure. First one
extracts the hadron masses and splittings from Euclidean propagators which are
then, in a second step, extrapolated and interpolated to the physical point in
the continuum limit. The first step is already discussed in \sec{se:mass}.

The second step of the analysis consists in performing the extra- and interpolations to
the physical point. As described in the previous sections, we have some
freedom in carrying out this procedure. First we can choose between
mass-independent or ratio method (see \sec{se:ratio}).  Second we can choose
different parameterizations for $\Delta M_X$ (see \sec{se:isofun}).  Third
there are two different minimum times for the fits of the relevant 
Euclidean correlators (see \sec{se:mass}).
Fourth we use two different scales $(8\tau)^{-1/2}=280~\mathrm{MeV}$ and
$(8\tau)^{-1/2}=525~\mathrm{MeV}$ to extract the renormalized electromagnetic
coupling (see \sec{se:qflow}).  Altogether, we obtain about
$\mathcal{O}(500)$ fits which are combined into a distribution
using the Akaike's information
criterion AIC (see \sec{se:aic}). The systematic error is obtained from the
variance of the distribution of different fit possibilities. Its mean gives
the central value.
The final
numbers and systematic errors of Table 1 and Figure 2 of the main text are
obtained using this procedure. Since estimating the systematic uncertainty
is a non-trivial task, we repeated the analysis using instead of the AIC weights the fit-qualities as a
weight or no weights at all. 
In all but one case, the results were within 
$0.2\sigma$ of the central values in Table 1 of the main text
For $\Delta\Xi_{cc}$ without using any weights the difference 
was  $0.7\sigma$.

The complete procedure was repeated on 2000 bootstrap samples, which were then
used to determine the statistical error.

\begin{figure}[H]
\centering
\includegraphics{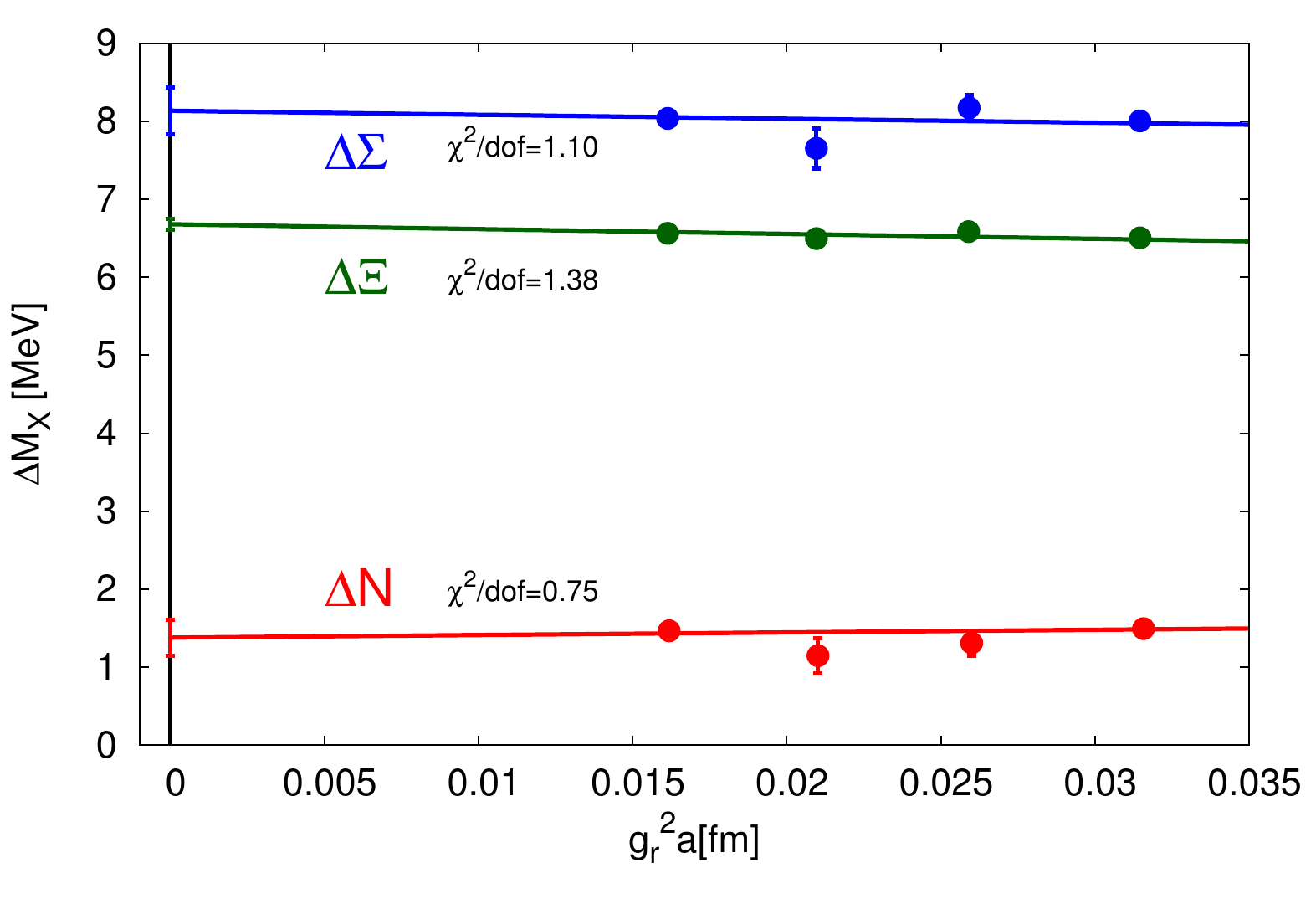}
\caption{\label{fi:mpidep}
    Lattice spacing dependence of the baryon octet isospin splittings in one
particular fit.
}
\end{figure}

In \fig{fi:mpidep} we show the three mass differences of the light hadrons in
one particular fit as the function of $g_r^2(a)a$, where $g_r(a)$ is the four-loop
running coupling \cite{vanRitbergen:1997va} at the scale of the lattice spacing.
The points are the results of the different ensembles averaged over a
common lattice spacing, which were transformed to the physical point in all
parameters except the lattice spacing using the fit function.  The line is the
result of the fit. 

In addition, three other, independent analyses were performed by different team
members to ensure a robust determination of the results and the error
estimation.  
We do not give the details of these procedures, we just mention that 
differences were e.g. one fits directly propagator ratios instead of 
the propagators themselves, jackknife is used instead of bootstrap 
to estimate statistical errors, 
one uses $w_0$~\cite{Borsanyi:2012zs} instead of the 
$\Omega$ mass to set the scale, one assumes the validity of the 
Coleman-Glashow relation instead of extracting it from our results, 
one uses a statistical test based on the binomial distribution instead of 
the Kolmogorov-Smirnov criterion to decide on the correlator fit 
time-ranges, the different models are weighed uniformly instead of 
using the fit quality (like AIC), etc. 
One of the procedures was a completely blind analysis: 
the extracted mass differences were multiplied by a random number 
between 0.7 and 1.3, and the analysis was carried out by a group 
member who did not know the value of this transformation factor. 
At the end of his analysis the transformation factor was removed. 
Comparing the results of these procedures with the main one, we found 
complete agreement.

\bibliography{main}

\newpage
\newpage

\newpage

\newpage

\newpage

\newpage

\newpage

\newpage

\newpage

\newpage

\newpage

\newpage

\newpage

\begin{table}[H]
\begin{center}
\begin{tabular}{|c|c|c|c|}
\hline
e&analytical \cite{Mertens:1997wx}&numerical&difference\\
\hline
0.21              & 0.191391 & 0.191378(05) & 0.000013(05) \\
$\sqrt{4\pi/137}$ & 0.201230 & 0.201155(10) & 0.000075(10) \\
0.40              & 0.215411 & 0.215161(10) & 0.000250(10) \\
\hline
\end{tabular}
\end{center}
\caption{\label{ta:qednum} Comparison of analytical and infinite volume extrapolated numerical results.}
\end{table}
\newpage

\begin{table}[H]
\begin{center}
\begin{tabular}{|c|c|c|}
\hline
$6/g^2$ & $a[\text{fm}]$ & $am_c$ \\
\hline
3.2 & 0.102 & 0.71 \\
3.3 & 0.089 & 0.58 \\
3.4 & 0.077 & 0.47 \\
3.5 & 0.064 & 0.35 \\
\hline
\end{tabular}
\end{center}
\caption{\label{ta:spa} Lattice spacing and charm quark mass at the four gauge couplings used in this work.}
\end{table}
\newpage

\begin{table}[H]
\begin{center}
\begin{tabular}{|c|c|c|c|c||c|c|c|}
\hline
$6/g^2$ & $am_u$ & $am_d$ & $am_s$ & $L^3\times T$ & $m_\pi[\text{MeV}]$ & $m_\pi L$ & \parbox{2cm}{\begin{center}$\times 1000$\\ trajectories\end{center}} \\
\hline
3.2 & -0.0686 & -0.0674 & -0.068 & $32^3\times 64$ & 405 & 6.9 & 1\\
3.2 & -0.0737 & -0.0723 & -0.058 & $32^3\times 64$ & 347 & 5.9 & 4\\
3.2 & -0.0733 & -0.0727 & -0.058 & $32^3\times 64$ & 345 & 5.8 & 1\\
3.2 & -0.0776 & -0.0764 & -0.05 & $32^3\times 64$ & 289 & 4.9 & 4\\
3.2 & -0.0805 & -0.0795 & -0.044 & $32^3\times 64$ & 235 & 4.0 & 12\\
3.2 & -0.0806 & -0.0794 & -0.033 & $32^3\times 64$ & 256 & 4.4 & 12\\
3.2 & -0.0686 & -0.0674 & -0.02 & $32^3\times 64$ & 440 & 8.1 & 4\\
3.2 & -0.0737 & -0.0723 & -0.025 & $32^3\times 64$ & 377 & 6.8 & 4\\
3.2 & -0.0776 & -0.0764 & -0.029 & $32^3\times 64$ & 317 & 5.6 & 4\\
3.2 & -0.077 & -0.0643 & -0.0297 & $32^3\times 64$ & 404 & 7.3 & 4\\
3.2 & -0.073 & -0.0629 & -0.0351 & $32^3\times 64$ & 435 & 7.8 & 4\\
3.2 & -0.077 & -0.0669 & -0.0391 & $32^3\times 64$ & 378 & 6.7 & 4\\
\hline
3.3 & -0.0486 & -0.0474 & -0.048 & $32^3\times 64$ & 407 & 6.1 & 1\\
3.3 & -0.0537 & -0.0523 & -0.038 & $32^3\times 64$ & 341 & 5.1 & 2\\
3.3 & -0.0535 & -0.0525 & -0.038 & $32^3\times 64$ & 340 & 5.0 & 2\\
3.3 & -0.0576 & -0.0564 & -0.03 & $32^3\times 64$ & 269 & 4.0 & 12\\
3.3 & -0.0576 & -0.0564 & -0.019 & $32^3\times 64$ & 281 & 4.2 & 12\\
3.3 & -0.0606 & -0.0594 & -0.024 & $48^3\times 64$ & 197 & 4.3 & 20\\
\hline
3.4 & -0.034 & -0.033 & -0.0335 & $32^3\times 64$ & 403 & 5.0 & 4\\
3.4 & -0.0385 & -0.0375 & -0.0245 & $32^3\times 64$ & 318 & 4.0 & 4\\
3.4 & -0.0423 & -0.0417 & -0.0165 & $48^3\times 64$ & 219 & 4.1 & 4\\
\hline
3.5 & -0.0218 & -0.0212 & -0.0215 & $32^3\times 64$ & 420 & 4.4 & 4\\
3.5 & -0.0254 & -0.0246 & -0.0145 & $48^3\times 64$ & 341 & 5.4 & 4\\
3.5 & -0.0268 & -0.0262 & -0.0115 & $48^3\times 64$ & 307 & 4.8 & 8\\
3.5 & -0.0269 & -0.0261 & -0.0031 & $48^3\times 64$ & 306 & 4.9 & 8\\
3.5 & -0.0285 & -0.0275 & -0.0085 & $48^3\times 64$ & 262 & 4.1 & 8\\
3.5 & -0.0302 & -0.0294 & -0.0049 & $64^3\times 96$ & 197 & 4.1 & 4\\
\hline
\end{tabular}
\end{center}
\caption{\label{ta:neu} List of ``neutral ensembles''.}
\end{table}
\newpage

\begin{sidewaystable}[H]
\begin{center}
\begin{tabular}{|c|c|c|c|c|c||c|c|c|}
\hline
$6/g^2$ & $e$ & $am_u$ & $am_d$ & $am_s$ & $L^3\times T$ & $m_\pi[\text{MeV}]$ & $m_\pi L$ & \parbox{2cm}{\begin{center}$\times 1000$\\ trajectories\end{center}} \\
\hline
3.2 & 1.00& -0.0819 & -0.0752 & -0.0352 & $32^3\times 64$ & 373 & 6.6 & 4\\
3.2 & $\sqrt{4\pi/137}$& -0.07788 & -0.07722 & -0.05022 & $32^3\times 64$ & 290 & 4.9 & 4\\
3.2 & 1.00& -0.0859 & -0.0792 & -0.0522 & $32^3\times 64$ & 290 & 4.9 & 4\\
3.2 & 1.41& -0.0943 & -0.0812 & -0.0542 & $32^3\times 64$ & 290 & 4.9 & 4\\
3.2 & 0.71& -0.0815 & -0.0781 & -0.0511 & $32^3\times 64$ & 290 & 4.9 & 4\\
3.2 & 1.00& -0.0889 & -0.0822 & -0.0462 & $32^3\times 64$ & 236 & 4.0 & 4\\
3.2 & 1.00& -0.0859 & -0.0792 & -0.0522 & $24^3\times 48$ & 292 & 3.7 & 5\\
3.2 & 1.00& -0.0859 & -0.0792 & -0.0522 & $48^3\times 96$ & 290 & 7.3 & 4\\
3.2 & 1.00& -0.0859 & -0.0792 & -0.0522 & $80^3\times 64$ & 289 & 12.2 & 1\\
\hline
3.3 & 1.00& -0.063 & -0.0555 & -0.0405 & $48^3\times 96$ & 335 & 7.4 & 4\\
3.3 & 1.00& -0.0666 & -0.0592 & -0.0329 & $48^3\times 96$ & 270 & 6.0 & 4\\
\hline
3.5 & 1.00& -0.034 & -0.02575 & -0.02575 & $32^3\times 64$ &411 &4.3 & 4\\
3.5 & 1.00& -0.0359& -0.0277 & -0.0173 & $48^3\times 96$ & 362  &5.7 & 4\\
3.5 & 1.00& -0.0389& -0.0307 & -0.0111 & $48^3\times 96$ & 283  &4.5 & 4\\
\hline
\end{tabular}
\end{center}
\caption{\label{ta:cha} List of ``charged ensembles''.}
\end{sidewaystable}
\newpage

\begin{table}[H]
\begin{center}
\begin{tabular}{|c|c|}
    \hline
    &$t_{min}[\fm]$ \\
\hline
$\Delta M_N$          & 1.1 \\
$\Delta M_\Sigma$     & 1.1 \\
$\Delta M_\Xi$        & 1.3 \\
$\Delta M_D$          & 1.1 \\
$\Delta M_{\Xi cc}$   & 1.2 \\
\hline
\end{tabular}
\end{center}
\caption{\label{ta:tmin} Starting time of the fit-intervals for different mass-splittings.}
\end{table}

\end{document}